\documentclass[12pt,preprint]{aastex}

\def\gtsim{\raise 2pt \hbox {$>$} \kern-1.1em \lower 4pt \hbox {$\sim$}}
\def\ltsim{\raise 2pt \hbox {$<$} \kern-1.1em \lower 4pt \hbox {$\sim$}} 

\shorttitle{A sample of low redshift BL Lacs. I.}
\shortauthors{Giroletti et al.}

\begin{document}

\title{A sample of low redshift BL Lacs. I. The radio data.}

\author{M. Giroletti and G. Giovannini}
  \affil{Istituto di Radioastronomia, via Gobetti 101, 40129, Bologna, Italy}
  \affil{Dipartimento di Astronomia, Universit\`a di Bologna, via Ranzani 1, 40127 Bologna, Italy}
\author{G. B. Taylor}
  \affil{National Radio Astronomy Observatory, P.O. Box O, Socorro, NM 87801, USA}
\and
\author{R. Falomo}
  \affil{Osservatorio Astronomico di Padova, vicolo Osservatorio 5, Padova, Italy}




\begin{abstract}

We present a new sample of 30 nearby ($z < 0.2$) BL Lacs, selected to
study the nuclear as well as the large scale properties of low power
radio sources. In this first paper, we show and discuss new radio data
taken with the VLA (19 objects at 1.4 GHz, either in A or C
configuration, or both) as well as with the VLBA (15 sources at 5
GHz).  On the kiloparsec scale, all objects exhibit a compact core and
a variety of radio morphologies (jets, halos, secondary compact
components). On the parsec scale, we find weak cores and a few short,
one-sided, jets.  From the jet/counter-jet ratio, core dominance, and
synchrotron self Compton model we estimate the intrinsic orientation
and velocity of the jets. The resulting properties of BL Lacs are
similar to those of a parent population composed of FR I radio
galaxies.

\end{abstract}

\section{INTRODUCTION}

Ever since the first suggestion of a possible unification of radio
sources \citep[e.g.,][]{orr82}, the viability for a comprehensive view
of Active Galactic Nuclei (AGN) has been actively debated. A major
achievement is the formulation of a scheme in which BL Lac objects and
quasars are the aligned counterparts of edge darkened (FR I) and edge
brightened (FR II) radio galaxies, respectively
\citep[e.g.,][]{urr95}. However, many important issues still need to
be worked out. In particular, the class of BL Lac objects and its
unification to FR I radio galaxies is an incessant source of puzzling
questions.

For many years, when only a few objects were known, an apparent
dichotomy was present within the BL Lac population. This dichotomy was
based initially on selection criteria, separating X-ray and radio
selected objects.  Later on, the dichotomy was refined according to
the position of the low energy peak in the spectral energy
distribution (SED), with BL Lacs classified as either low (LBL) or
high (HBL) frequency peaked objects. LBLs have the peaks of emitted
energy ($\nu F_{\nu}$) in the infrared/optical and MeV/GeV bands while
HBLs peak at higher frequencies, namely UV/X-ray and GeV/TeV;
furthermore, other observational differences between LBL and HBL have
been found, including the fact that HBL have less dominant radio
cores, a lower degree of optical polarization, and a possibly
different cosmological evolution \citep[see][and references
therein]{rec03}. However, multi-wavelength studies of larger samples
seem to have solved this dichotomy, showing that the SED of BL Lacs
(and emission line blazars) form a continuum, and that the emission
peaks shift to higher frequency as the bolometric luminosity decreases
\citep[the so called {\it blazar sequence},
see][]{fos98}. \citet{don01} confirmed this trend considering more
sources and hard X$-$ray data. Among possible explanations,
\citet{ghi98} proposed that an increase in luminosity causes the
emitting particles to suffer more severe radiative losses; in turn,
this explains the shift of the peaks to lower frequency. There are
however some problems with this scenario: following the discovery of a
significant number of FSRQs with flat $\alpha_{RX}$ in a deep survey
\citep[the DXRBS, see][]{per98,lan01}, \citet{pad03} considered a
large sample of $\sim 500$ blazars, finding no evidence of
anticorrelations between $\nu_{\rm peak}$ and radio power (or other
related quantities), as expected from the blazar sequence. Moreover,
\citet{cac04} present a few examples of low radio power BL Lacs with
steep $\alpha_{RX}$, suggesting that there may exist sources with a
peak at low frequency and steep $\alpha_{RX}$, as well as objects
peaking at high frequency but still displaying flat $\alpha_{RX}$
\citep[see also][]{ant04}.

In spite of our advances in understanding, most physical parameters
remain difficult to determine in BL Lac objects. Several papers
\citep[see e.g.][and references therein]{fal02} have addressed the
long debated issue of determining the mass of the central black hole,
finding values of order $10^8 - 10^9 M_\odot$. The situation remains
more complicated for the jet physics and main parameters. On one hand,
we have constraints from the detection of very energetic GeV and TeV
photons, which require Doppler factor values $\delta > 10$
\citep{tav98}; on the other hand, lower values are preferred from
radio observations and from the lack of measured superluminal motions
in most TeV sources \citep{gir04,pin04,tin02}. Even in sources where
proper motions have been found \citep{hom01,jor01}, the estimated jet
velocity from radio data is lower than the jet velocity required by
high energy photons. This discrepancy suggests that the $\gamma$-rays
and the radio photons may be produced in different regions of the jet,
moving at different bulk velocities or differently oriented with
respect to the line of sight. Furthermore, it is also possible that
the jet has a dual {\it transverse} velocity or energetic structure;
this may be needed to explain the different behaviour of optical cores
in BL Lacs with respect to FR I radio galaxies \citep{chi00}. In
particular, the optical radiation could come from a faster spine,
while the radio emission could originate in a slower, external layer
of the jet.  Evidence for a velocity structure in parsec scale jets
has been found in some sources as 1144+35, Mkn 501, 3C\,264, M87,
0331+39, 1055+018 \citep[see
e.g.][]{sol89,han96,gio01,gir04,per99,per01,att99}.  The presence of a
velocity structure could be due to the jet interaction with the
interstellar medium \citep{gio01} or could be an intrinsic jet
property \citep{mei03}.

As a matter of fact, the radio interferometry technique is an unique
tool to determine parameters of the jet in the region from a few
parsecs out to the kiloparsec scale structure, by means of studying
observational properties such as jet sidedness, core dominance, and
proper motion of knots \citep{gio03}. Very Long Baseline
Interferometry (VLBI) studies have yielded significant results for a
number of sources, including a possible signature of the double
velocity structure in the limb brightened jet of Markarian 501
\citep{gir04}. However, systematic and detailed VLBI studies on
sizeable samples of BL Lacs are mostly based on bright, flux-limited
catalogues, such as the 1 Jy sample \citep{sti91}. By its very
definition, this sample consists mostly of powerful and distant
objects. These bright sources have yielded some interesting results,
among which is a high incidence of FR II structures in the parent
population \citep{rec01,cas99}. Moreover, objects in this sample
present often a significant bending between parsec and kiloparsec
scale structure \citep{cas02}, resulting in a peculiar bimodal
distribution of the misalignment angle between the structure on the
two scales \citep[see also][and references therein]{app96}.

As for the weaker objects, parsec scale studies similar to those
discussed above are still missing or based on small numbers of objects
\citep{rec03,kol96}. Given the SED shape for BL Lacs, objects weak in
the radio are usually classified as HBL. Thus far, almost all BL Lacs
detected at TeV energies belong to this sub-class. Moreover, weak
objects are more easily found at low redshift, so that the high
angular resolution of VLBI techniques allows observers to investigate
the very innermost regions ($1 \ \mbox{mas} \sim 2 \ \mbox{pc}$ at
$z=0.1$)\footnote{Throughout this paper, we assume $H_0 = 71 \,
\mbox{km} \, \mbox{s}^{-1} \, \mbox{Mpc}^{-1}$ and $q_0=0.5$.}.

In order to investigate the above issues and to extend the current
view to the low-$z$ objects, we concentrated our attention on
the sample of \citet{fal00}, who studied the host galaxies of 30 low
redshift ($z<0.2$) BL Lacs with the {\it Hubble Space Telescope}
(HST). We carried out new radio observations of this sample, to look
for differences in morphology with more powerful objects and to derive
parameters for jet physics. In this paper, we present the new
observations, and summarize data from the literature.  By using all
available data we can study the intrinsic power and parent
population. In a forthcoming paper (Giroletti et al. in prep.), we will
focus on the radio/optical comparison.

The sample selection is illustrated in \S \ref{sec:sample}, along with
the main results of \cite{fal00}. Our observations are described in \S
\ref{sec:observations}; we present the results in \S \ref{sec:results}
and discuss them in \S \ref{sec:discussion}. Finally, \S
\ref{sec:conclusions} summarizes the main results.

\section{THE SAMPLE}
\label{sec:sample}

\subsection{Sample selection}

By merging 7 flux limited samples\footnote{The 7 samples are: 1 Jy
\citep{sti91}; S4 \citep{sti94}; PG \citep{gre86}; HEAO-A2
\citep{pic82}; HEAO-A3 \citep{rem99}; EMSS \citep{mor91,sto91,rec00}; Slew
\citep{sch93,per96}}, \citet{sca00a} have collected a total of 132 BL
Lac objects. The HST snapshot image survey of BL Lacs
\citep{urr00,sca00a} has provided a homogeneous set of short-exposure,
high-resolution images through the F702W filter for 110 objects in the
sample. From this large dataset, \citet{fal00} extracted a sub-sample
of 30 low redshift ($z< 0.2$) objects for which it was possible to
perform a detailed study of the properties of the host galaxy.

We now discard three objects from this sample: 1853+671 and 2326+174,
which have a redshift slightly larger than 0.2 ($z=0.212$ and 0.213,
respectively), and 2005$-$489, which is too far south to be observed with
comparable quality high resolution imaging. Conversely, we add three more objects, which also have $z<0.2$ and high quality HST observations available. The objects are 1215+303 ($z=0.130$, Perlman et al. private communication), 2254+074 ($z=0.190$), and 1652+398 (Mkn 501, $z= 0.034$). In total we have 30 objects. Although the sample cannot be considered complete, the selection process appears free from significant biases with respect to source orientation and jet velocity.

\begin{figure}
\plotone{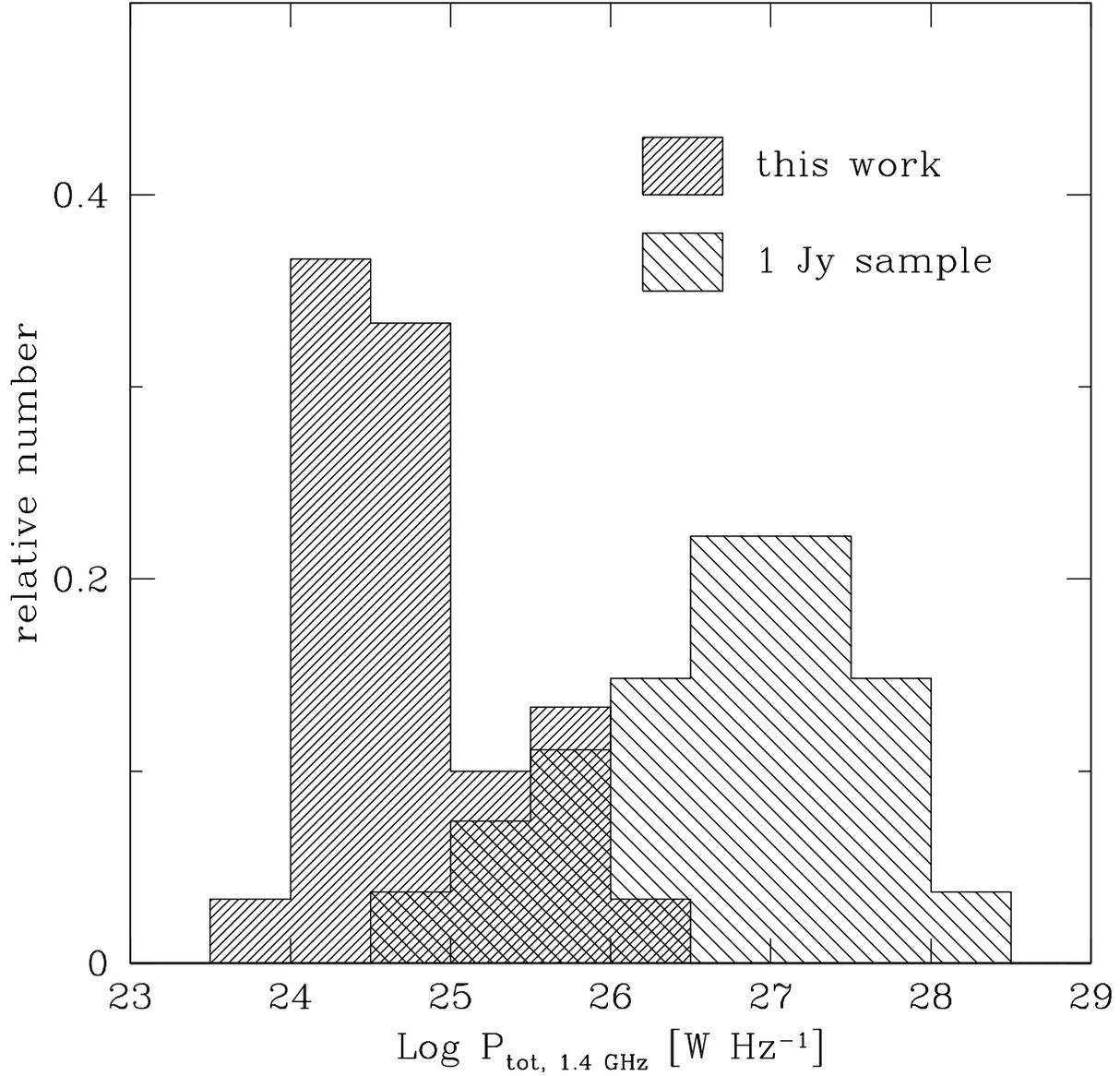} 
\figcaption{Histogram of total power distribution at 1.4 GHz for
objects in our sample and in the 1 Jy sample. Data from the NVSS
survey \citep{con98} for our sample and from \citet{rec01} for the 1
Jy sample.
\label{fig1} }
\end{figure}

Table \ref{table1} presents the full list of objects in the
sample\footnote{The optical spectra of 0521$-$365 \citep{sca95} and
2201+044 \citep{ver93,fal94} show emission and absorption lines and
are similar to Seyfert 1 spectra but with lines of lower
luminosity. At HST resolution the optical sources are fully resolved
into nucleus, host galaxy, and a jet that contains some
structures.}. HBLs are clearly dominant, and most objects have been
originally selected at X-ray energies. This selection is significantly
different from most previous studies at radio frequencies. We note
that HBLs have lower total luminosities and are particularly weak in
the radio. Figure \ref{fig1} plots, as histograms, the distribution of
total power at 1.4 GHz for objects in the present sample compared to
the 1 Jy sample. The two distributions are significantly different
with $\langle \mathrm{Log} P_{1.4 {\rm GHz}} \rangle = 24.7 \pm 0.6 \,
{\rm W} \, {\rm Hz}^{-1}$ and $26.7 \pm 0.9 \, {\rm W} \, {\rm
Hz}^{-1}$, respectively (some objects in the 1 Jy sample lack redshift
information and were not considered). This is largely due to the large
flux limit of the 1 Jy sample; furthermore, with no cut in redshift,
it contains a number of high-$z$ BL Lacs that bias the sample toward
more powerful objects.

However, we note that there is some overlap between the two
populations. Six of the objects belong to both samples (1418+546,
1514$-$241, 1652+398, 1807+698, 2200+420, 2254+074); indeed, these are
well known objects, such as BL Lac itself, Mkn 501, and 3C\,371. In
contrast, most of the weakest HBL lack radio images on the parsec
and/or kiloparsec scale.

\subsection{Host Galaxies: a Summary}

For these low-$z$ objects it was possible with HST to investigate
features and structures in the host galaxy that are undetectable with
ground-based observations \citep{fal00}. The host galaxy can be
investigated within less than 1~kpc from the nucleus (corresponding to
0''.3 at $z = 0.2$).  In particular one can search for slight offsets
of the nucleus with respect to the galaxy center, determine the
presence of sub-components in the host galaxy, investigate the
detailed shape of the isophotes, and search for dust features and
close companions.

The detailed analysis of HST images indicates that, in spite of the
presence of the active nucleus, the host galaxy appears in most cases
to be a completely ``normal'' elliptical.  Both the ellipticity and
the isophotal shape distributions are similar to those for radio
galaxies and radio-quiet ellipticals. This suggests that tidal
interactions are very infrequent or are short-lived with respect to
the nuclear activity time scale.

There are no indications of displacements and/or off-centering of the
galaxy isophotes with respect to the nucleus, meaning the unresolved
nuclear source truly sits in the center of the galaxy. This rules out
the microlensing hypothesis for BL Lacs, which predicts frequent
offsets for the nucleus with respect to the host galaxy.

The HST images have consistently shown that BL Lac hosts are virtually
all massive ellipticals (average luminosity M$_R = -23.7 \pm 0.6$ and
effective radius $R_e$ = 10 kpc) typically located in poor groups.
Moreover, these data have shown that no significant difference is
present between the hosts of HBL and those of LBL in spite of the
remarkable differences in their SED \citep{urr00}.

\section{OBSERVATIONS AND DATA REDUCTION}
\label{sec:observations}

\begin{figure}
\plotone{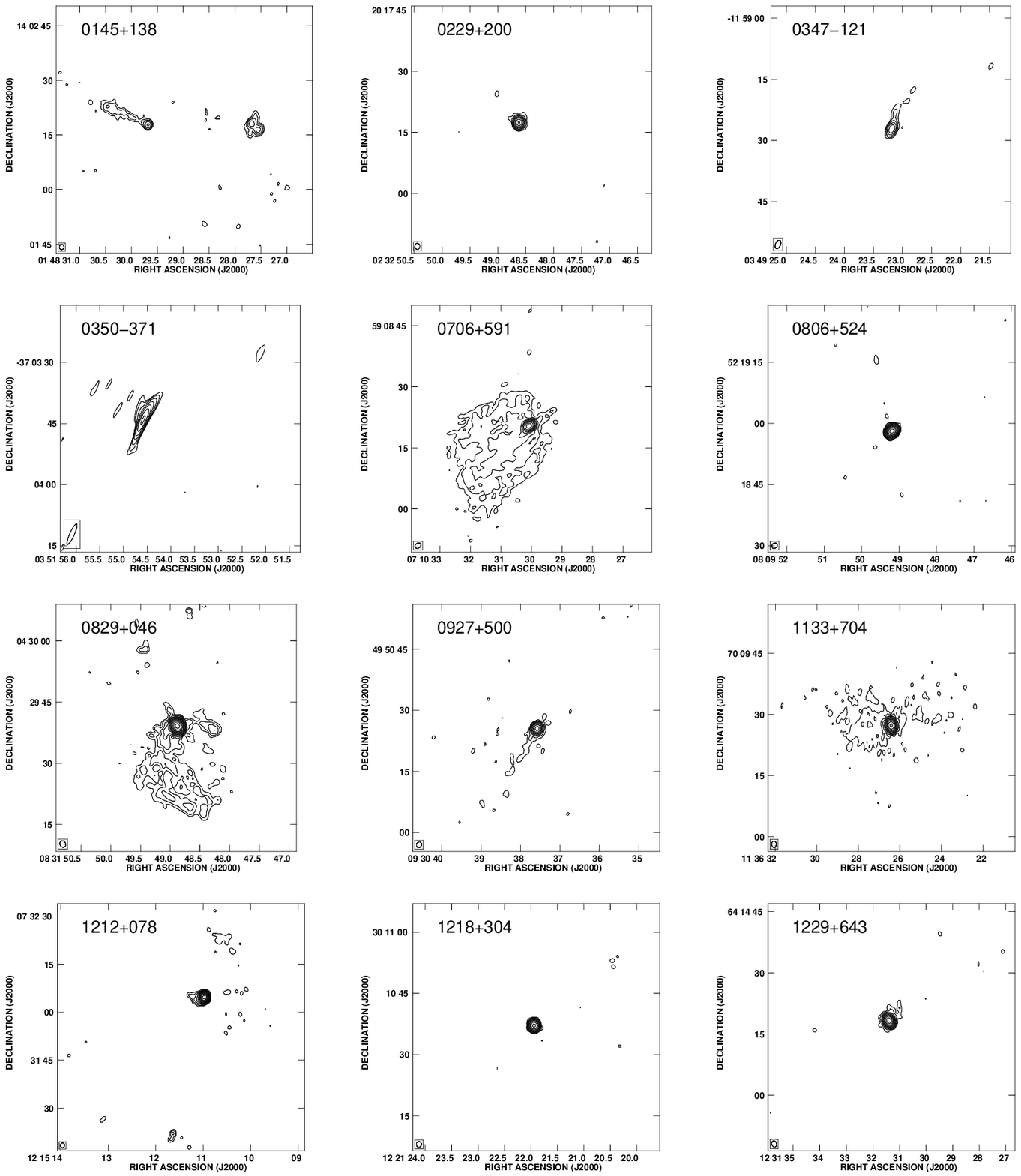}
\figcaption{VLA images taken in A configuration page 1 of 2. Contours are drawn at (1, 2, 4, 8, 16, ...) times the noise level. Noise levels and image peaks are given in Tab. \ref{table4}.
\label{fig2} }
\end{figure}

\begin{figure}
\epsscale{1}
\plotone{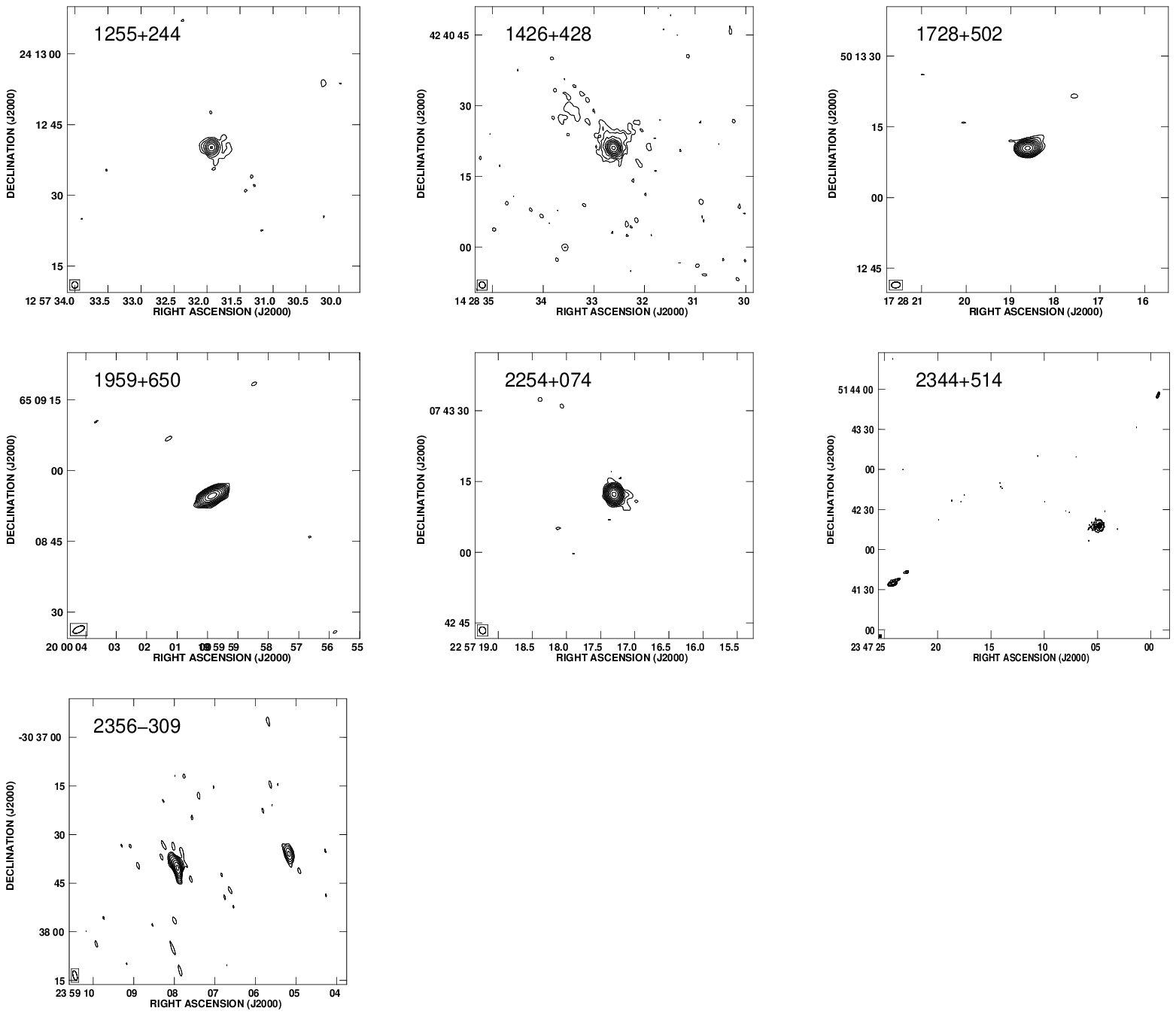}
\figcaption{VLA images taken in A configuration continued. Contours
  are drawn at (1, 2, 4, 8, 16, ...) times the noise level. Noise
  levels and image peaks are given in Tab. \ref{table4}.
\label{fig3} }
\end{figure}

\begin{figure}
\plotone{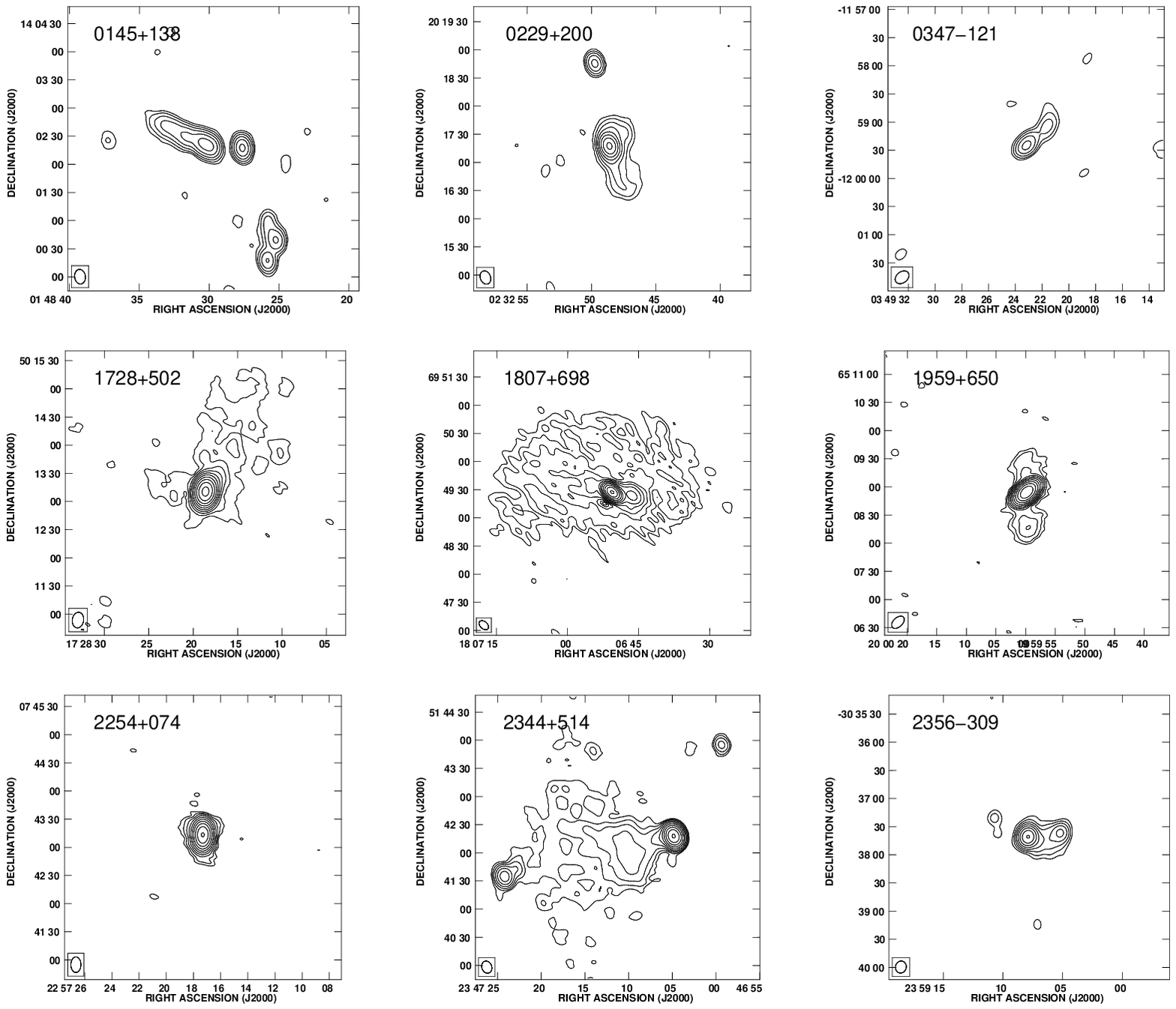}
\figcaption{VLA images taken in C configuration. Contours are drawn at (1,
  2, 4, 8, 16, ...) times the noise level. Noise levels and image peaks
  are given in Tab. \ref{table4}.
\label{fig4} }
\end{figure}

\begin{figure}
\plotone{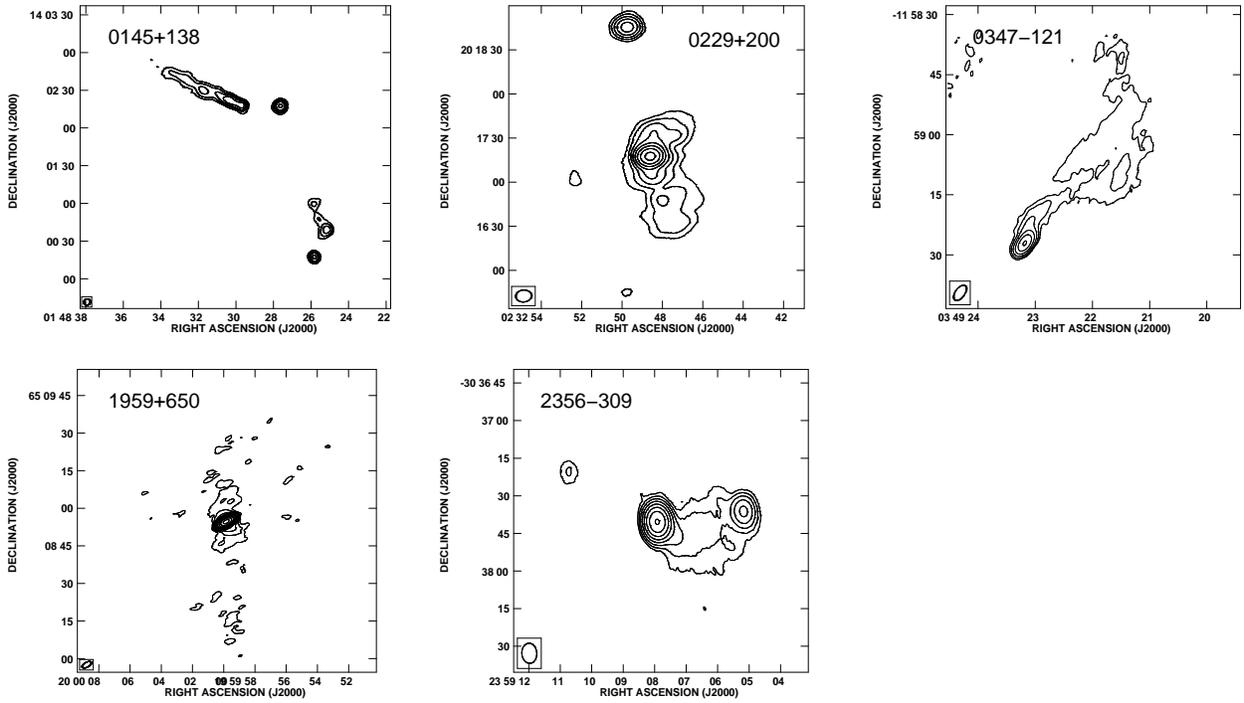}
\figcaption{VLA images made from a combination of data from the A and
C configurations.
\label{fig5} }
\end{figure}

\begin{figure}
\epsscale{.9}
\plotone{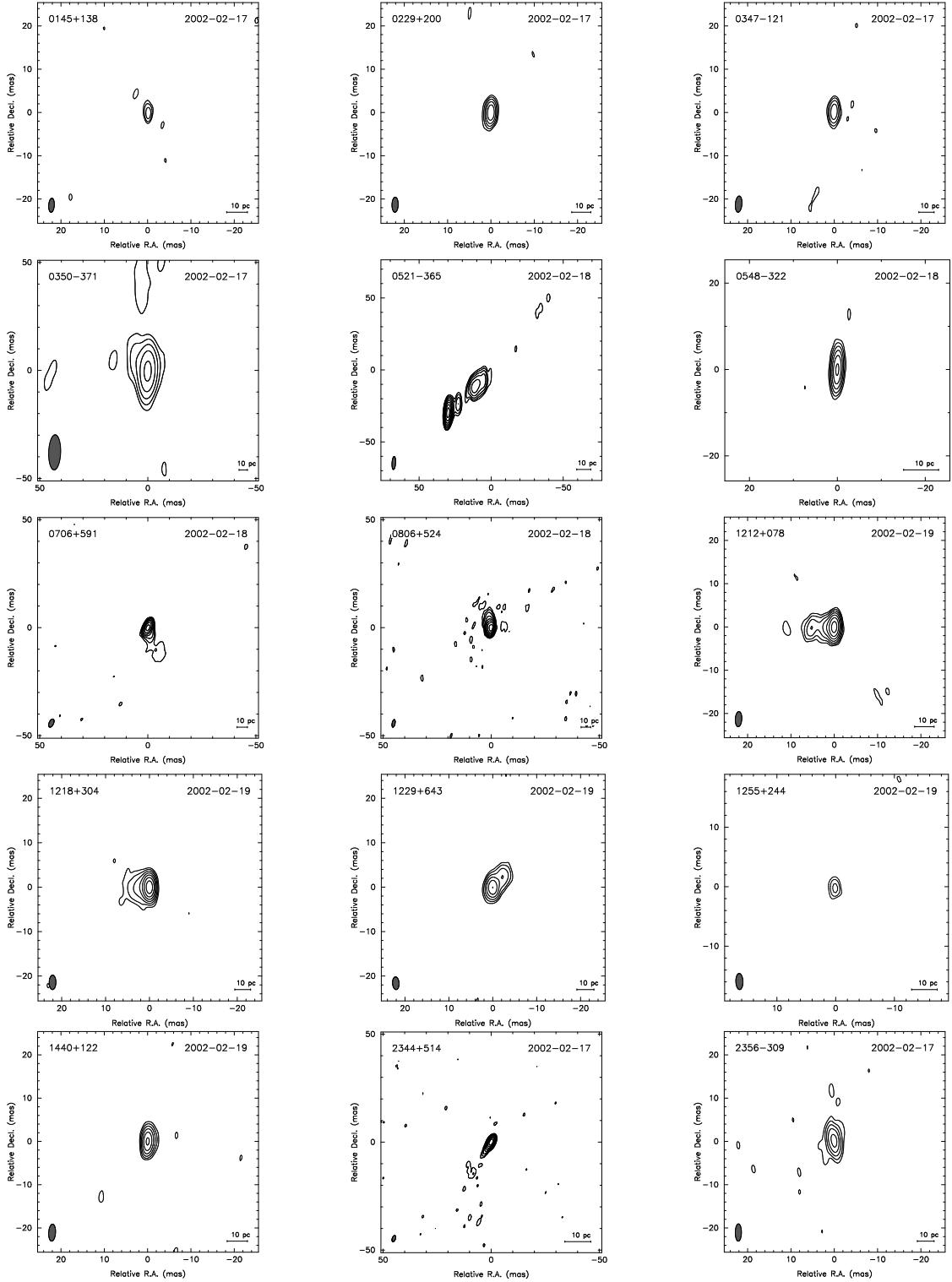}
\figcaption{VLBA images. Contours are drawn at (1, 2, 4, 8, 16, ...)
  times the noise level. Noise level and image peaks are given in
  Tab. \ref{table4}.
\label{fig6} }
\end{figure}

Observations were aimed at completing the imaging for all the sources
in the sample, both on arcsecond (Table \ref{table2}) and
milliarcsecond (Table \ref{table3}) scales. We present the final
images in Fig. \ref{fig2}--\ref{fig6} (see Tab. \ref{table4} for image
parameters). Objects with good images available in the literature were
not re-observed; see Table \ref{table5} for a summary of references.

\subsection{VLA observations}

We obtained observations with the NRAO\footnote{The National Radio
Astronomy Observatory is operated by Associated Universities, Inc.,
under cooperative agreement with the National Science Foundation.}
Very Large Array (VLA) at 1.4 GHz on 2002 February 22 and May 3 in A
configuration and on 2002 October 8 in C configuration (see Table
\ref{table2} for a summary of the observations). In order to optimize
the $(u,v)$ coverage we observed in snapshot mode with two or three
scans per source at different hour angles. On average, each source was
observed for about 16 minutes after allowing for observations of
calibrators and slewing of the telescope. We used the NRAO
Astronomical Image Processing System (AIPS) to reduce the data and
perform imaging with the standard phase self-calibration
technique. Amplitude self-calibration was used only for the strongest
sources ($>$ 100 mJy) to obtain a better dynamic range. For fainter
sources, amplitude self calibration is not necessary since images are
noise limited and not dynamic range limited; moreover, spurious
solutions could be introduced because of the low signal to noise
ratio. We produced images with different weight distribution (uniform
and natural) and did not find significative differences. Images
presented here have been obtained with a weight in between the natural
and uniform ({\tt ROBUST} 0 in the task {\tt IMAGR} in AIPS). This
weighting scheme yields a typical final restoring beam of $\sim 1.9''
\times 1.2''$ for A configuration and of $\sim 15'' \times 11''$ for C
configuration data; the average noise is $\sim 70 \, \mu$Jy/beam and
$\sim 140 \, \mu$Jy/beam, respectively. Images of sources in the
southern hemisphere have more elliptical beams and slightly higher
noise.

In total, we imaged 19 objects with the array in A configuration and 9
in C configuration. For the eight sources observed with both arrays,
we combined the two final self-calibrated data-set using AIPS task {\tt DBCON}. This yields
better sensitivity (about a factor $\sqrt 2$) and intermediate
resolution. However, the $(u, v)$-coverage is not optimal, and the
process of self-calibration can fail to converge, especially when
extended flux detected in C configuration is over-resolved by the A
array. For this reason, we only present combined array images for five sources
(0145+138, 0229+200, 0347$-$121, 1959+650, and 2356$-$309).

\subsection{VLBA observations}

Very Long Baseline Interferometry observations have been performed
with the NRAO Very Long Baseline Array (VLBA) for 15 sources on 2002
February 17, 18, and 19, at a frequency of 5 GHz (see Table
\ref{table3} for a summary of the observations).  A single VLA antenna
was substituted for the VLBA antenna at Pie Town.  We recorded 4 IFs
(at 4971.49, 4979.49, 4987.49, 4995.49 MHz) with 8 MHz bandwidth each,
in full polarizations. The data were correlated in Socorro and the
reduction was performed in AIPS. All data were globally fringe fitted
\citep{sch83} and then self calibrated. However, the VLA observations
in A configuration provided us with new and precise position
measurements. Since we had good positions for our target sources, we
were able to avoid fringe-fitting the data for weak sources, which
easily fails.  Amplitude calibration was initially done using the
standard method of measuring system temperatures and antenna gains
calibration, followed by gain refinement using strong point-like
calibrators.

On average, we have 45 minutes of useful data for each source,
providing a nominal thermal noise of $\sim 0.11 \, \mbox{mJy} \,
\mbox{beam}^{-1}$. All our images have noise figures within a factor 2
from this. All sources are detected above the $3 \sigma$ level, even
the weakest.  For 6 sources, only a faint point-like component was
detected. For the remaining 9 strongest sources, we imported the $(u,
v)$ data into {\sc Difmap} \citep{she94,she95} for further
self-calibration and imaging.

\section{RESULTS}
\label{sec:results}

\subsection{arcsec scale}

We present our final images in Fig. \ref{fig2}--\ref{fig5} (see
Tab. \ref{table4} for image parameters) and a list of flux densities
for all objects in Tab. \ref{table5}. We also include low frequency
(325 MHz) data (Column 4), as measured in the Westerbork Northern Sky
Survey \citep[WENSS,][]{ren97} or from the Texas Survey
\citep{dou96}. For sources lacking flux density measurement from both
surveys, we estimate the flux density from the 1.4 GHz NRAO VLA Sky
Survey \citep[NVSS,][]{con98} assuming the average spectral index
obtained from other sources, i.e. $\alpha = 0.11$. This is a sensible
choice as all values obtained are below the sensitivity threshold (or
out of the sky coverage) of both surveys. For 0521$-$365 only, whose
radio data is missing because of its location, we interpolate the
large number of measured fluxes available in the literature from 80
MHz to 8800 MHz.  Columns 5 and 6 give the total and core flux
densities at 1.4 GHz, and Column 7 lists the ratio between these
values.  We also report the core flux density at 5 GHz in Col. 8 and
the corresponding reference in Col. 9.  Finally, in Col. 10 we
classify sources according to their arcsecond scale morphology.

For the sake of homogeneity, the total flux densities at 1.4 GHz
(Col. 5) are all measured on data from the NVSS. We used, however,
information from our higher resolution observations in order to
separate possible contribution from confusing sources: the A array
images show that this is needed only in 0145+138, in which we
subtracted 8 mJy from the NVSS total. For objects observed also in C
configuration, we consider the total flux density measured on our
images as well and find that the deviation from the NVSS datum is
always within $\pm 10 \%$; such a small difference in the total flux
density does not affect our results and it is probably due to core
variability effects. In any case, there are no indications that the
NVSS systematically overestimates the total flux density.
The core value (Col. 6) has been measured using {\tt JMFIT} in AIPS on
our A configuration images (or from the literature, as given in the
notes); the corresponding ratio between core and total flux is listed
in Col. 9.  In general, the core is the strongest component and
accounts for most of the flux. In two objects (1418+546 and
1514$-$241), the core flux density is even larger than the total; this
is probably due to variability of the core. At the other extreme,
there are a number of sources (10/30) where a significant fraction of
flux ($> 50 \%$) is found in extended regions or in secondary
components.

 We find 9 sources where the core is the dominant
component; 8 objects with a core-jet morphology; 4 with a core-halo
morphology; 7 objects with more than one compact component -- possibly
a hot spot or a background source; and 2 objects (0548$-$322 and
0829+047) which are located in rich clusters and are clearly
wide-angle-tail (WAT) radio sources.

The objects span a range in total power of about 2.5 orders of
magnitude at 1.4 GHz. The weakest object is 1255+244, which has
Log$P_{\rm t} = 23.88 \, {\rm W} \, {\rm Hz}^{-1}$, whilst the most
powerful is 0521$-$365, with Log$P_{\rm t} = 26.07 \, {\rm W} \, {\rm
Hz}^{-1}$.

\subsection{milliarcsecond structure}

On VLBI scales, our 5 GHz observations have a typical resolution of
$\sim 3.8 \, \mathrm{mas} \times 1.5$ mas and noise level of a few 0.1
mJy/beam. Figure \ref{fig6} presents the images; Table \ref{table6}
summarizes the most significant parameters: peak flux densities, total
intensity, presence of a jet, and jet position angle (PA).

All objects are dominated by a strong, compact component. In five
sources this component is responsible for all the correlated flux
density and in two cases (0548$-$322 and 1440+122) there is little
($\sim 10\%$) flux density in extended structures that we could not
image. Short one-sided jets (typically $< 10$ mas) are present in
seven sources and only 0521$-$365 presents a longer jet; in any case,
the core is always the strongest component. No counter-jet is detected
in any of the sources.

We have modelfitted the visibilities of all the sources in Difmap. In
most cases, we need only one component (in addition to the core) to
describe the jet. When two or more components are required, they
usually align on a straight path, without showing significant
bending. However, a comparison to the largest scale structure
indicates that some bending occurs at large distance from the core.

Finally, from our model-fit we determine the jet brightness $B_{\rm
j}$ and the jet/counter-jet ratio $R_{\rm min}$, which are presented
in Columns 7 and 9 of Table \ref{table6}. Since no object in the
sample has a detectable counter-jet, the values presented are only
lower limits based on the noise of the images. This is also true for
objects with values derived from literature data (see notes to Table
\ref{table6}).

\subsection{Parsec/Kiloparsec Scale Comparisons}

\begin{figure}
\plotone{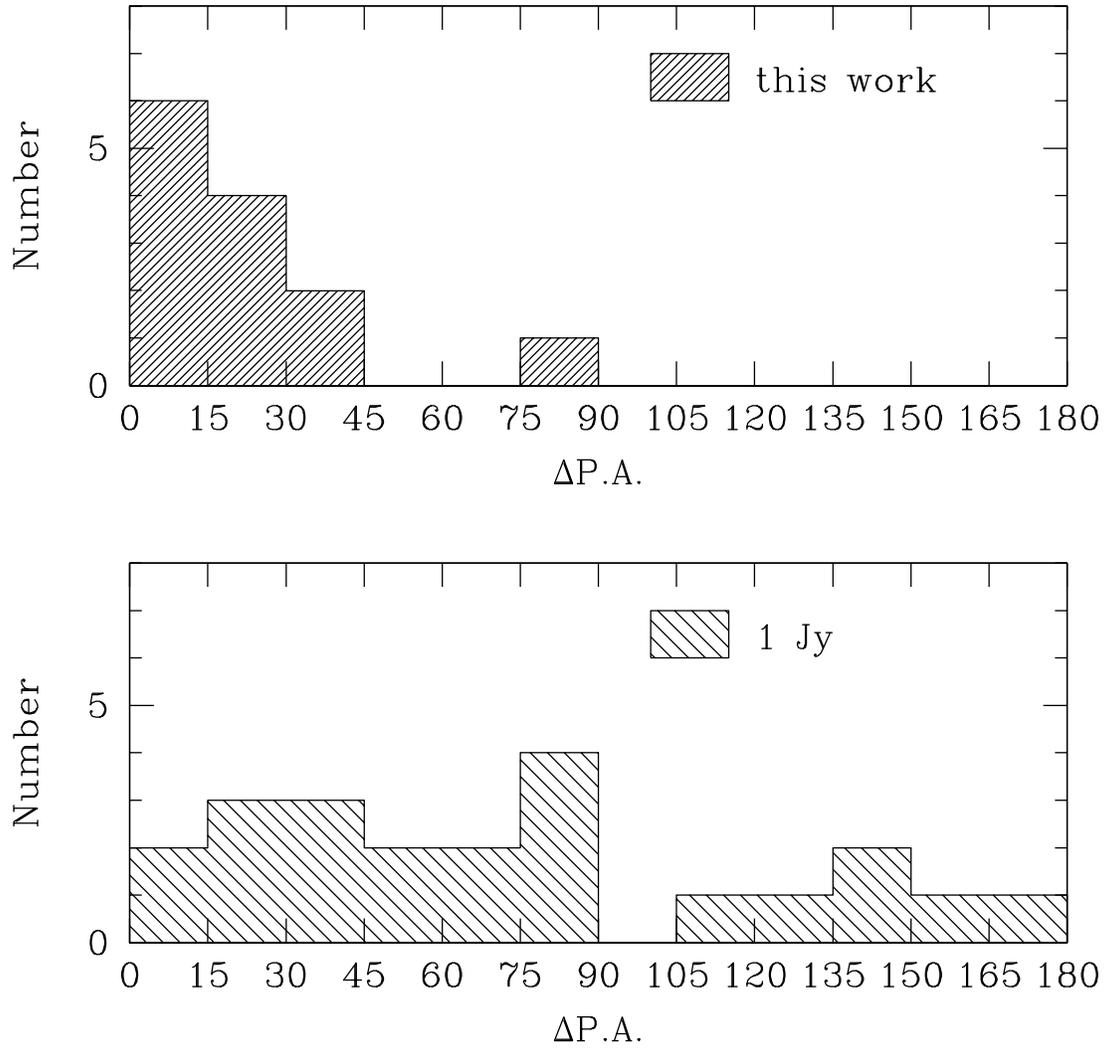}
\figcaption{Distribution of bending for sources in the present sample
  (top) and in the 1 Jy sample (bottom, see text). We
  have excluded from our sample objects belonging to the 1 Jy sample. 
\label{histo}}
\end{figure}

A comparison of the morphology on parsec and kiloparsec scales reveals
in general a good agreement, and objects in the present sample do not
exhibit large bending between small and large scales.  There are 10
sources that bend $\le 30^\circ$, two between $30^\circ$ and
$60^\circ$, and two more are $>60^\circ$; for the others there is no
preferred direction on either parsec or kiloparsec scale. Even with
this uncertainty, this result seems to support the view proposed by
\citet{rec03}, i.e. that HBLs tend to have less bent jets than LBLs
(see Fig. \ref{histo} and next section for discussion).  It is also
worth noting that all three objects showing an optical and X-ray jet
(0521$-$365, 1807+698, and 2201+044, all LBLs), present straight jets
as well. Therefore, with few exceptions, it seems that large bending
in BL Lacs is not common; this agrees with the expectations of unified
models, given the findings of \citet{gio01} that FR I radio galaxies
show a good agreement between parsec and kiloparsec scale jet
PA. Large bending in the jets is also unlikely in those objects
presenting a core+halo morphology, such as 1133+704, 1215+303
\citep[see FIRST image,][]{bec95}, 1426+428, 1807+698, and
2344+514. The classic explanation is that the halos are lobes seen end
on, and the main axis of the object is closely aligned with the
line-of-sight.  This explanation argues against large bends in the jet
in these sources as well.

In the comparison between the arcsecond core at 5 GHz and the
correlated VLBI flux at the same frequency (given in Tab. \ref{table5} and
\ref{table6}, respectively) there is a continuity of
behavior -- only 9 sources have integrated VLBI flux densities $\le 60
\%$ of their arcsecond core flux density. This result suggests that
only a small fraction ($\sim 30 \%$) of BL Lacs objects have a complex
sub-arcsecond structure not visible in our data. In the majority we
can follow the structure from parsec to kiloparsec scales.

Observational data suggest that beaming effects play a major role on
the parsec scale and become less important on larger (kpc) scales. One
clear piece of evidence is the presence of the symmetric structure on
the kiloparsec scale in sources with an asymmetric parsec scale
morphology.  In sources discussed here we have 5 sources where VLA
images suggest/show the presence of symmetric structure. They are:
0229+200, 0706+591, 0829+046, 1807+698, and 1959+650. These results
imply that at a projected distance of 5-10 kpc from the core the jet
velocity is no longer relativistic. On the contrary, some sources are
still asymmetric in the VLA images. They are: 0145+138, 0347$-$121,
0927+500, and 1212+078. Assuming intrinsically symmetric jets and
Doppler favoritism as the origin of observed asymmetries, the radio
jets of these sources are still relativistic on scales of tens of kpc.

\subsection{Notes on individual sources}

{\it 0145+138 --} This is one of the weakest objects (${\rm Log}
P_{\rm tot, 1.4 GHz} = 24.19 \, {\rm W} \, {\rm Hz}^{-1}$), to show a
core-jet structure. The jet is unusually long for such a low-power
object, as it extends for $\sim 200$ kpc eastward. On the opposite
side of the core, $\sim 25$" westward, there is a secondary component,
which is associated with an elliptical galaxy with the same redshift,
as previously noted by \citet{per96} \citep[see also][]{sli98}; our A
array image resolves this source into a double. Other nearby sources
are possibly related to the foreground galaxy cluster Abell 257. Note
that the total flux given in Tab. \ref{table5} refers to the BL Lac
source only; the contribution of the nearby source has been
subtracted. The VLBA image shows only a 3 mJy core in agreement with
the core flux density from the VLA in A configuration.  We surmise
that most of the flux density in this source is from the long,
one-sided jet.

{\it 0229+200 --} The C-array image suggests, and the A+C image
confirms, the presence of a two-sided jet, with the main jet pointing
to the south, in agreement with \citet{rec03}. About 100" to the
north, we find a compact radio source not yet identified
optically. The flux density of the VLBI core is $\sim 16$ mJy; no jet
is visible on the parsec scale. Thanks to a longer exposure and better
$(u,v)$-coverage, the VLBA image by \citet{rec03} shows a parsec scale
jet, well aligned with the kpc scale main jet.

{\it 0347$-$121 --} This object shows in the A array image a jet in PA
$-12^\circ$; at 7'' from the core the jet bends by $\sim 35^\circ$ and
expands with a large opening angle, showing in the A+C image a
lobe-like structure. The VLBI image gives no indication of a parsec
scale jet.

{\it 0350$-$371 --} Because of its low declination (${\rm Dec} =
-37^\circ$), this object has a somewhat lower resolution A
configuration image. There is an indication of a jet present both on
arcsec and milliarcsec scales, as an elongation of the main component
to the north-east in PA 33$^\circ$ and 46$^\circ$, respectively. This
yields a small bend of $\Delta {\rm PA} \sim - 13^\circ$ from the
parsec to the kiloparsec scale.

{\it 0521$-$365 --} This is a nearby ($z=0.055$), bright EGRET source
\citep{lin95}. The optical spectrum of this objects exhibits prominent
and variable emission lines \citep{sca95} and it was also classified
as N galaxy and Seyfert galaxy, although its host galaxy is a luminous
giant elliptical (Falomo 1994).  This source is also well known for
the presence of a prominent radio and optical jet
\citep{dan79,kee86,falomo94}, which resembles that of the nearby radio
galaxy M87 \citep[i.e.][]{spa94}. The optical jet is well aligned with
the kiloparsec radio jet, and the radio and optical structures have a
clear correspondence \citep{sca99}. Our VLBA image shows that the same
PA found on the parsec scale jet is maintained, without any
significant bending, over three orders of magnitude. This is
consistent with a relatively large angle of view, in agreement with
the absence of superluminal motion showed by \citet{tin02} and the
findings of \citet{pia96}.

{\it 0548$-$322 --} This southern object resides in a rich
environment, with close companions and other galaxies at the same
redshift \citep{fal95}. This is in agreement with the wide-angle tail
(WAT) structure revealed on kiloparsec scales by \citet{lau93} and
\citet{rei99}. The parsec scale image is dominated by a 35 mJy core,
with little or no other emission detected. The large difference
between this value and the kiloparsec total flux density suggests that
the emission on large scales is spread over an extended low brightness
area.

{\it 0706+591 --} Our image in A configuration is a significant
improvement in resolution with respect to the only image published so
far \citep{lau93}. The main structure is quite extended, broad, and
the core is located at the north-west edge of this quite round
cocoon. It could be a tailed radio structure (WAT or NAT) viewed at a
small angle of sight. Note however that there is no indication of an
overdensity of galaxies in its vicinity. A small orientation angle is
also suggested by the different direction of the parsec scale core-jet
structure ($\Delta{\rm PA} \sim 80^\circ$).

{\it 0806+524 --} This source is heavily core-dominated. It looks
point-like with the VLA, with a flux density of 160 mJy/beam at 1.4
GHz. On the VLBI scale, the integrated flux density at 5 GHz is 137
mJy, distributed in a core and a short ($\sim 5$ mas), northbound (PA
$\sim 13^\circ$) jet.

{\it 0829+046 --} This is one of the rare BL Lacs in which there is
evidence of emission on both sides of the core. In the VLA A array
image, two symmetric jets emerge at PA $110^\circ$ (and
$-70^\circ$). Both jets bend, showing a radio structure typical of
head-tail radio galaxies. For this reason, we believe that the large
bending of the jet is related to this motion and not intrinsic to the
source \citep[see also][]{ant85}. This leaves us with a bend between
parsec and kiloparsec scale of 44$^\circ$ \citep[see VLBA image
in][]{jor01,fey00,lis98}.

{\it 0927+500 --} This is the second most distant source in the sample
and looks like a faint ($\sim 20$ mJy), core-dominated source. No
information is available on the parsec scale morphology.

{\it 1133+704 --} Our observation reveals a core-halo morphology,
confirming the claim of \citet{war84}. The halo is faint and heavily
resolved in the image in A configuration. This explains the results of
\citet{lau93}: in their 5 GHz image, the halo is resolved out and only
the compact 125 mJy core is detected. The major axis of the halo is in
the east-west direction, i.e. well aligned with the parsec scale jet
detected by \citet{kol96}.
 
{\it 1212+078 --} This object presents a good alignment between the
parsec and kiloparsec scale structure. The jet is oriented at $\sim
90^\circ$ and is detected for 12 mas with the VLBA and almost 50" with
the VLA. \citet{rec03} present a VLBA image (1997 May 17) and a deep B
array VLA image suggesting a transverse orientation for the extended
structure. A compact feature 40 mas south of the core is present in
our image as well as in the one of \citet{rec03}. It is likely that
our VLA image resolves some of the extended emission, which is also
visible in \citet{per96} on the opposite side, as well as in the FIRST
and NVSS images.

{\it 1218+304 --} We confirm that this source is compact on kiloparsec
scales \citep[see][and references therein]{per94,lau93} and place an
upper limit on its size of 0.19"; at the redshift of 1218+304
($z=0.182$, Perlman et al., private communication), this corresponds
to 0.7 kpc. The VLBA image reveals a 10 mas jet emerging at PA $\sim
90^\circ$. The total VLBA correlated flux is 57 mJy and dominates the
total flux density of the source.

{\it 1229+643 --} The kiloparsec scale structure shows an almost
unresolved core \citep[see also][]{per94}, with a little emission from
a halo-like feature. However, even if the poor signal-to-noise of this
structure does not allow us to make any strong claim about its nature,
we can estimate that it is oriented at a PA of $\sim -20^\circ$. This
is in general agreement with the previous image from
\citet{per93}. The VLBA jet points in direction north-west ($\mbox{PA}
= - 43^{\circ}$) and the core-jet structure accounts for most of the
flux density from this source. Therefore, we tentatively propose a
$\Delta {\rm PA}$ \ltsim $25^\circ$.

{\it 1255+244 --} This extremely weak object is only marginally
resolved with the VLA, and barely detected by the VLBA ($\sim 3$ mJy
core).  This is the weakest object in the sample ($\mathrm{Log}P_{\rm tot, 1.4
GHz} = 23.5 \, {\rm W} \, {\rm Hz}^{-1}$ )

{\it 1426+428 --} The comparison of the NVSS total flux density (61
mJy) and the peak of the A array image (32 mJy/beam) indicates that
some extended emission is present.  Our image confirms the presence of
a faint halo surrounding the central core, but does not recover all
the flux. Since previous images in C array \citep{lau93} detected an
intermediate value of 46 mJy, we believe that the difference has to be
ascribed to resolution effects rather than variability.  The NW
extension is oriented at ${\rm PA} \sim 50^\circ$ and therefore
presents a $\Delta {\rm PA} = 30^\circ$ with respect to the inner
structure studied by \citet{kol96}. Note that this is a TeV source
\citep{aha02,hor02}.

{\it 1440+122 --} Little information can be obtained for this
object. The VLBA image reveals only a faint (15 mJy) core, possibly
extended to the west. VLA images obtained by \citet{gio04} show a
nuclear flat spectrum nuclear emission surrounded by a steep spectrum
halo structure $\sim$ 8" in size at 8.4 GHz moderately elongated in
the SW direction.

{\it 1728+502 --} The 200 mJy core dominates the source; it is
slightly resolved to the north-west in the A configuration image.  The
C array VLA image detects $\sim 16$ mJy of extended emission in the
same direction (in PA $-30^\circ$), spread over more than 100". A
bending of $\sim 25^\circ$ west is required from the parsec scale jet
shown by VLBA \citep{kol96} and EVN+MERLIN images (Giroletti et al.,
in preparation).

{\it 1807+698 --} This is a well known source (3C 371), with a jet
detected in the optical with the HST \citep{sca99} and in the X-rays
with {\it Chandra} \citep{pes01}. This jet does not bend significantly
and is well aligned (in ${\rm PA} \sim 100^\circ$) with the parsec
scale jet imaged with the Space VLBI by \citet{gom00}. The orientation
is in good agreement with the PA of the jet visible in our C array
image ($\Delta {\rm PA}$ consistent with 0) and of the main lobe
detected in the deep B array image at 5 GHz by \citet{wro90}; see also
the 5 GHz and 15 GHz VLA observations of \citet{ode88}. Thanks to the
lower frequency and more compact configuration, we also detect a
diffuse halo of about 340 mJy, surrounding both lobes and extending
over 200\arcsec \citep[see also][]{cas99}.

{\it 1959+650 --} \citet{cos02} suggested that this object could be a
TeV source; in fact, a strong detection of very high energy
$\gamma$-rays was obtained with the Whipple 10 m telescope
\citep{hol03} and the HEGRA Cherenkov telescopes \citep{aha03},
confirming the preliminary result presented by \citet{nis00}. Our C
array radio image clearly shows a peculiar two-sided structure. The
combination of A and C data indicates that the symmetry could be in
the jet region, suggesting that the source is oriented in the plane of
the sky and/or that the jet is non-relativistic on the arcsecond
scale. The parsec scale jet in the image by \citet{rec03} at 5 GHz has
a PA of $-5^\circ$ indicating that the jet does not change direction
over three orders of magnitude. Note, however, that high frequency (15
GHz) observations suggest a 1 mas jet to the southeast, in PA $\sim
160^\circ$\citep{pin04}.

{\it 2254+074 --} This source has one of the brightest cores (${\rm
Log} P_{\rm c, 5 GHz} = 26.1 \, {\rm W} \, {\rm Hz}^{-1}$). Only a
little extended emission is present on arcsecond scale, both in the A
and C configuration. A comparison to the parsec scale total flux density
\citep[350 mJy, ][]{fey00}, indicates that some emission originates
on intermediate scales.

{\it 2344+514 --} This is one of the most puzzling sources in the
sample. VLA images show two bright components within 200'' (180 kpc)
from the position of the optical and radio core. Furthermore,
extended, low-brightness emission is present between the core and the
eastern feature. The A array observation resolves the faint emission,
revealing a core-halo morphology, while the eastern feature appears
extended in the direction of the core. Since this extension is visible
also in the C array image \citep[as well as in the B array map
of][]{rec03}, it cannot be ascribed to bandwidth smearing. This
elongation and the diffuse radio bridge connecting it to the core
indicates that the component is related to the source.  On the
contrary, the NW component does not show any radio structure
suggesting a connection. Finally, the VLBA image shows a jet oriented
at 142$^\circ$ \citep[see also][]{rec03}, confirming the complexity of
this source ($\Delta {\rm PA} \sim 45^\circ$ to the VLA main
axis). Notice also that \citet{cat98} report a TeV detection from this
source and that {\it Chandra} observations reveal diffuse X-ray
emission in its environment \citep{don03}.

{\it 2356$-$309 --} Two components separated by 36'' are connected by a
weaker ``bridge'', resolved by the VLA in A configuration. A
third component is found on the opposite side. The 
core is barely detected on the parsec scale, due to the low
declination ($-30^\circ$) and flux density (21 mJy) of this BL
Lac. Because of the presence of a core, a radio bridge, and two
possibly related compact structures, this source is similar to
2344+514.

\section{DISCUSSION}
\label{sec:discussion}

According to the current view of the unification of AGNs, BL Lac
objects are expected to be FR~I radio galaxies oriented at a small
angle with respect to the line-of-sight. Moreover, radio data
\citep[e.g.,][]{gio01} as well as variability, one sidedness,
superluminal motion and high frequency emission (X- and $\gamma$-
rays) strongly support the existence of relativistic jets. In
agreement with the presence of fast jets and small angles to the line
of sight, BL Lacs are usually core-dominated objects. However, we note
that in our sample we have 10/30 objects where the core flux density
is $< 50\%$ of the total flux density at 1.4 GHz.

Under the standard assumption that jets are intrinsically symmetric,
we can derive constraints on the jet velocity and orientation.  The
approaching jet is amplified by relativistic beaming and the
counter-jet is de-boosted.  The observed ratio $R$ between the
observed jet and counter-jet brightness is related to the intrinsic
jet velocity $\beta$ and orientation angle $\theta$ by $R =
\displaystyle \left(\frac{1+\beta\cos\theta}{1-\beta
\cos\theta}\right)^{p}$, where $p = 2+\alpha$ (continuous jet) or $p =
3+\alpha$ (moving sphere). Here we assume a continuous jet, with
spectral index $\alpha = 0.5$ ($S (\nu) \propto \nu^{\ - \alpha}$),
following \citet{gio94} and references therein.

However, in BL Lacs the counter-jet is generally de-boosted below the
detection threshold on the parsec scale and we can only determine
lower limits $R_{\rm min}$. Furthermore, since our sample is dominated
by faint objects, the main jet shows in most cases a low brightness
emission resulting in a small $R_{\rm min}$. Therefore, in most cases
we do not find significant constraints on $\beta$ and $\theta$; only
in 6 objects the J/CJ ratio is $R_{\rm min}>100$. In these cases, we
give the relative limits on $\beta \cos \theta$ in Table \ref{table6}
(Col. 10).

The relativistic boosting at the base of the jet affects the observed
radio power of the core; by comparing this value to that expected from
the total power at low frequency via the correlation found by
\citet{gio88,gio01}, we can derive the amount of boosting. We account
for variability, which results in a range of possible Doppler factors
for each source after allowing for a variation of a factor 2 in the
core flux density. Considering these intervals, we estimate for each
object the allowed orientation angle under the assumption of a Lorentz
factor $\Gamma = 5$. We note that larger Lorentz factors do not change
significantly the permitted angles, while values $\Gamma < 3$ are not
allowed, according to \citet{gio01} and from strong and widely
accepted lines of evidence (superluminal motions, rapid variability,
high observed brightness temperatures, detection of $\gamma$-ray
emission).

The Synchrotron Self-Compton model \citep[SSC,][]{ghi93} poses a third
observational constraint on the Doppler factor, by requiring that the
X-ray flux produced by Inverse Compton does not exceed that
observed. Thanks to the availability of a large number of observations
carried out by X-ray satellites ({\it Einstein}, EXOSAT, {\it ROSAT},
ASCA, and {\it Beppo}SAX), \citet[][see references therein]{don01}
have collected fluxes at 1 keV for all the objects in the sample. The
standard SSC formula \citep{mar87} yields the Doppler factor as an
inverse power of the VLBI core angular dimension; since VLBI cores are
unresolved in our data, we find lower limits on $\delta$. There are 20
sources where the result is not informative; however, for 11 sources
we have mild limits on $\delta$, as reported in Table
\ref{table7}. These limits provide useful constraints on the minimum
velocity and the maximum angle allowed. Basically, these values are in
good agreement with those required from the core enhancement argument;
given the independence of the two estimates, this strengthens the
conclusion that small to moderate angles to the line-of-sight and high
velocity jets are characteristic of the objects in the sample.

\begin{figure}
\epsscale{1}
\plotone{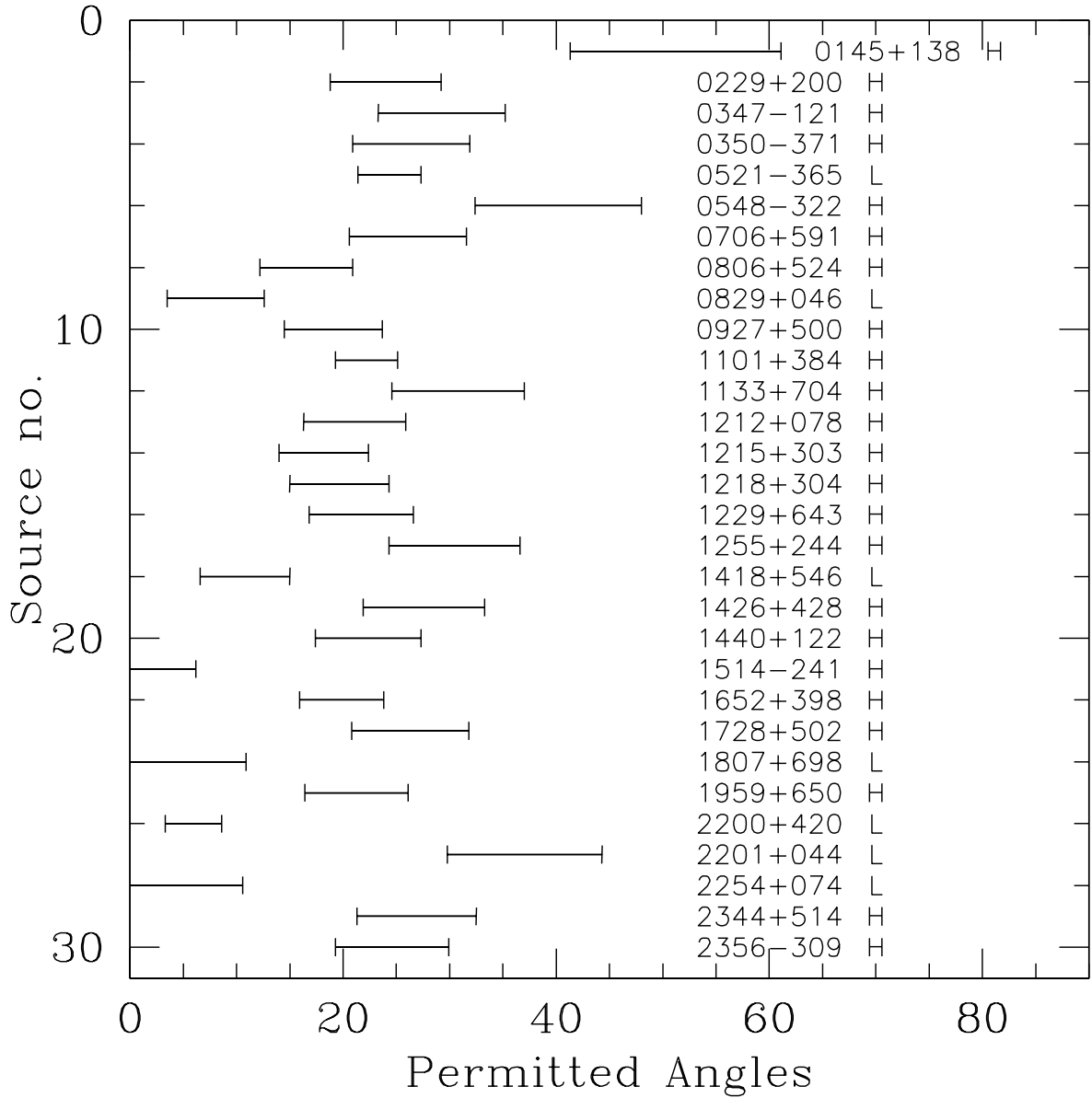}
\figcaption{Allowed angles to the line-of-sight, assuming $\Gamma =
  5$. The letter beside the name refers to the type of source (H: high
  frequency peaked BL Lac; L: low frequency peaked BL Lac). 
\label{angles} }
\end{figure}

We illustrate the resulting allowed angles in Fig. \ref{angles}. The
derived angles are consistent with the unified scheme, since most of
the BL Lacs in our sample are oriented at a small to moderate angle to
the line of sight. In Table \ref{table8}, we present the resulting
best available estimates from the previous studies of the allowed jet
orientation angle $\theta$ and the relative Doppler factor $\delta$ in
the case $\Gamma = 5$. We note that, despite the low number of LBLs
considered here, they seem to be oriented at smaller angles ($\langle
\theta \rangle_{\rm LBL} = 14^\circ \pm 12^\circ$) with respect to
HBLs ($\langle \theta \rangle_{\rm HBL} = 25^\circ \pm 9^\circ$). A
Kolmogorov-Smirnov test on the viewing angles for the two populations
of HBL and LBL yields a probability $< 8 \times 10^{-3}$ that the data
sets are drawn from the same distribution (K-S statistic
$d=1.66$). Notice however the rather large uncertainties, which are
due to the small number of objects but also to a likely large scatter
in the distribution of intrinsic angles. 

\begin{figure}
\plottwo{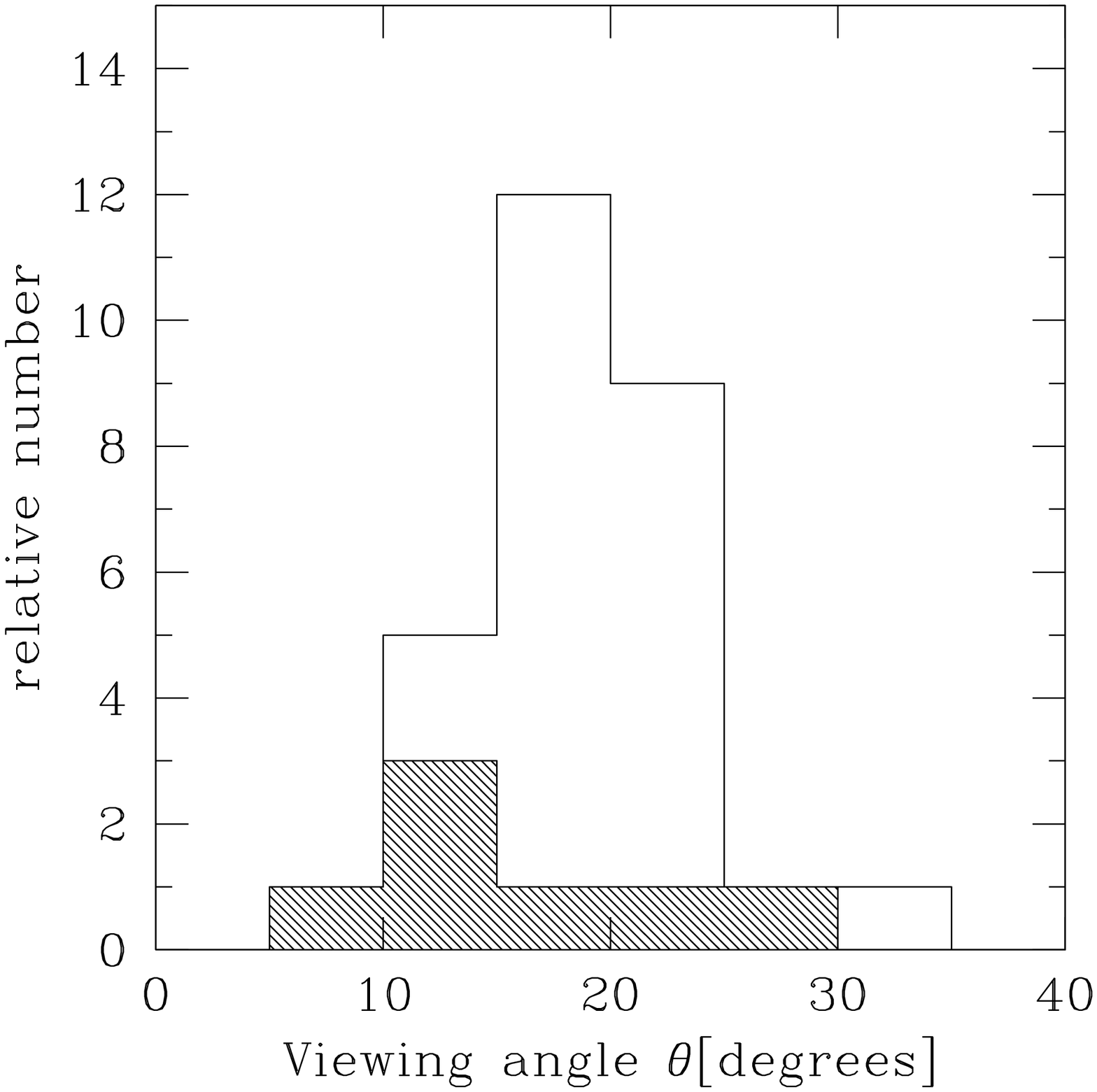}{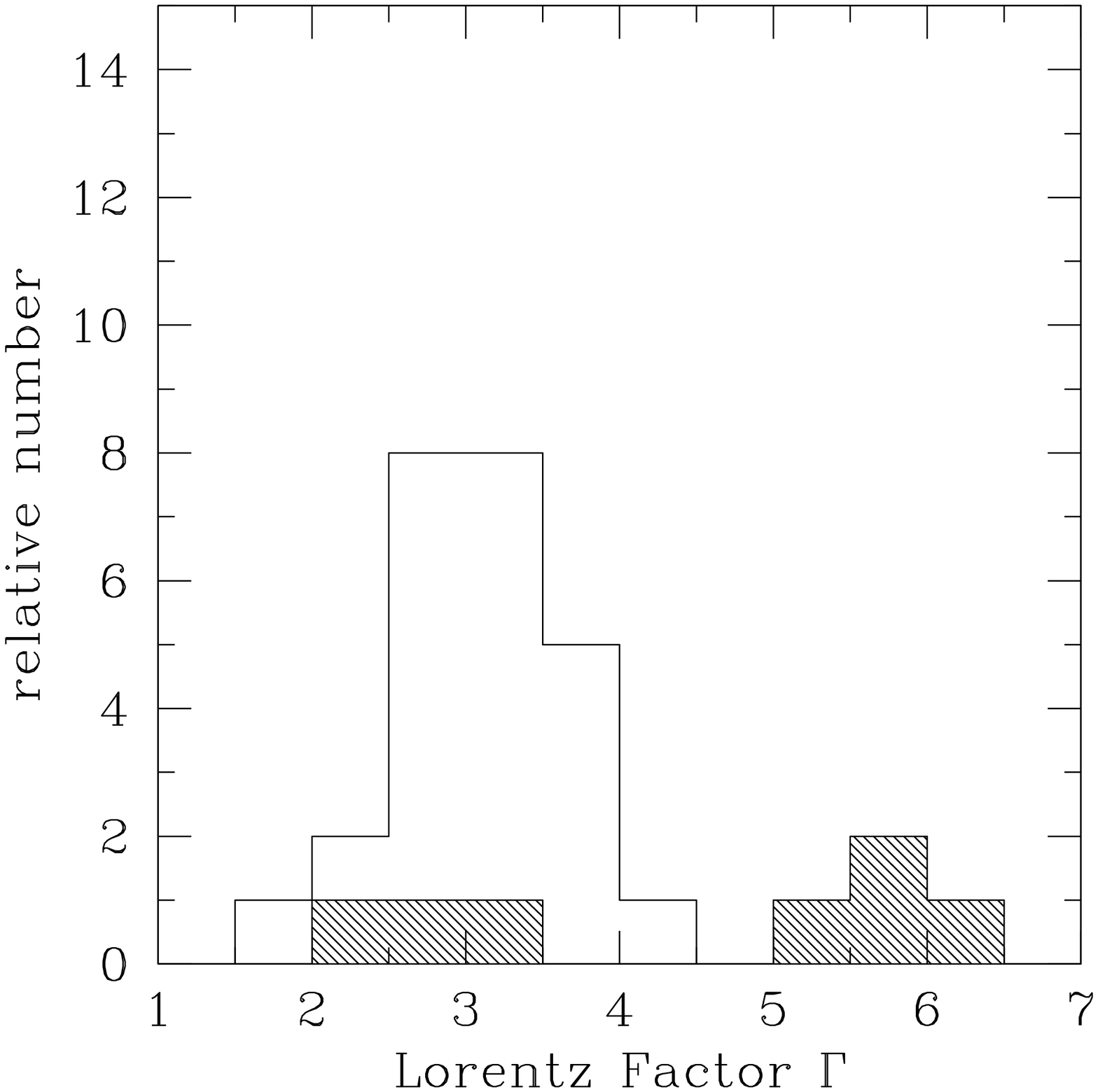}
\figcaption{Distribution of the resulting viewing angle $\theta$
  (left) and Lorentz factor $\Gamma$ (right), assuming $\Gamma \sim
1/\theta$; the shaded parts
  correspond to LBL only.
\label{histos} }
\end{figure}

The adoption of a single Lorentz factor can also be partly responsible
for this result. Although the assumption of $\Gamma = 5$ is a good
approximation of the real situation, we explore another approach, less
restrictive than assuming a constant velocity for all the jets.  In
particular, we consider the relation $\Gamma \sim 1/\theta$, which
corresponds to the maximum angle allowed for a given Doppler factor
and is, in a statistical sense, the most likely situation. We still
make use of the $P_\mathrm{c}/P_\mathrm{t}$ relation, in order to
uniquely determine the values of $\beta$ and $\theta$. We show in
Fig.~\ref{histos} the histogram of the resulting angles and Lorentz
factor distribution. The average angle for the sample is $\langle
\theta \rangle = 18^\circ \pm 5^\circ$, without showing any relevant
difference between LBL and HBL ($\langle \theta_{\rm LBL} \rangle =
15^\circ \pm 7^\circ$ and $\langle \theta_{\rm HBL} \rangle = 20^\circ
\pm 4^\circ$, respectively). There are still four LBL that need to be
oriented at rather small angles ($\theta < 15^\circ$), namely 0829+046
($z=0.180$), 1418+546 ($z=0.152$), 2200+420 ($z=0.070$), and 2254+074
($z = 0.190$); however, their redshifts are typically larger than the
average, so these may be more extreme and peculiar objects. In
Fig. \ref{histos} (right panel), we also show the distribution of
Lorentz factors: the majority of objects, including all HBLs, are
distributed around $\Gamma = 3$. There are however four objects
separated from the others, which have $\Gamma > 5$; these objects are
the four LBLs mentioned above.

Depending on the assumptions made on the jet bulk velocity, we are
left with two alternatives on the beaming properties of the jet:
either our HBLs are viewed at somewhat larger angles than LBL, or the
bulk velocity of the radio jets for the HBL population is
intrinsically smaller. The former explanation has been put forward
several times in the literature \citep[e.g.][]{cel93,jan94}, although
it cannot satisfactorily account for {\it all} observed properties of
BL Lacs \citep{sam96,rec01}. Conversely, the latter interpretation has
been less explored and is puzzling, since constraints on smaller
scales, and from other arguments (e.g. TeV $\gamma-$ray emission),
require an opposite behaviour. However, we note that, on parsec
scales, the study of proper motions have revealed superluminal
components in EGRET sources \citep{jor01} and subluminal or absent
motion in TeV sources \citep{tin02,pin04}. Thus, it is interesting to
speculate that the emission of TeV photons taking place on even
smaller scales may be responsible for energetic losses resulting in
slower jets on radio scale. Another possible explanation for the
decrease in jet velocity invokes the properties of the interstellar
medium (ISM), which can vary among galaxies. Note that evidence of a
strong jet deceleration within $\sim$ 5 kpc from the core has been
found in 3C\,449 \citep{fer99}, 3C\,31 \citep{lai02}, M\,87
\citep{bir99}, and 3C\,264 \citep{bau97,lar04}.

\begin{figure}
\plotone{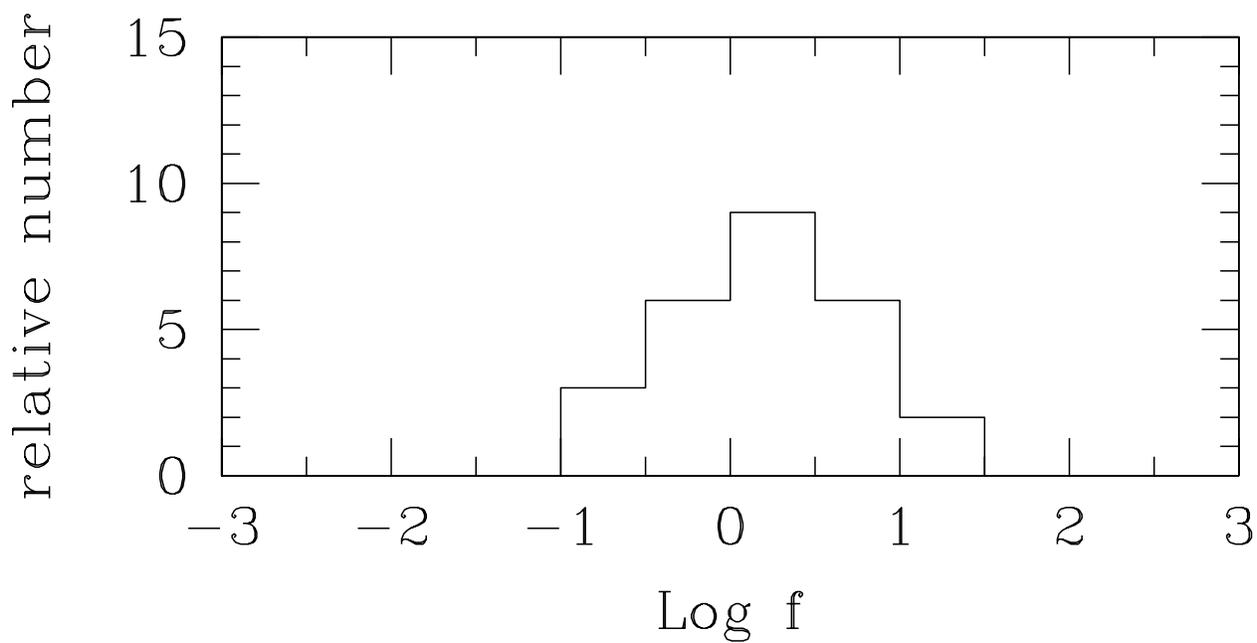}
\figcaption{Distribution of the core dominance factor $f$ for sources in the present sample.
\label{fig:erre}}
\end{figure}

Since samples of distant objects with large flux limits (e.g. the 1
Jy) have been studied in detail, it is worthwhile to compare some
properties of the two populations. The value of the core dominance
parameter $\displaystyle f = \frac{S_{\rm core}}{S_{\rm ext}}$ in our
sample is $\langle f \rangle = 3.2$ (see
Fig. \ref{fig:erre}). Actually, we excluded the three sources
(1418+546, 1514$-$241, and 2254+074) in which the flux density of the
core is larger than the total; this behaviour has to be ascribed to
variability, and reminds us that this result needs to be considered
with caution. However, this is true also for other samples and we do
not expect it to affect the average properties of the sources in our
sample. The present low redshift BL Lac sample is less core-dominated
than the EMSS and the 1 Jy; in the EMSS $\langle f \rangle \ge 4.2$
\citep{rec00} and much larger values are observed in the bright 1 Jy
BL Lacs \citep{rec01}.  Correspondingly, the core is frequently
associated with a resolved radio morphology, including halos,
secondary components, and symmetric two-sided jets; this also suggests
that kiloparsec scale jets may have different properties, being either
non-relativistic or still mildly relativistic in different sources.

\begin{figure}
\plotone{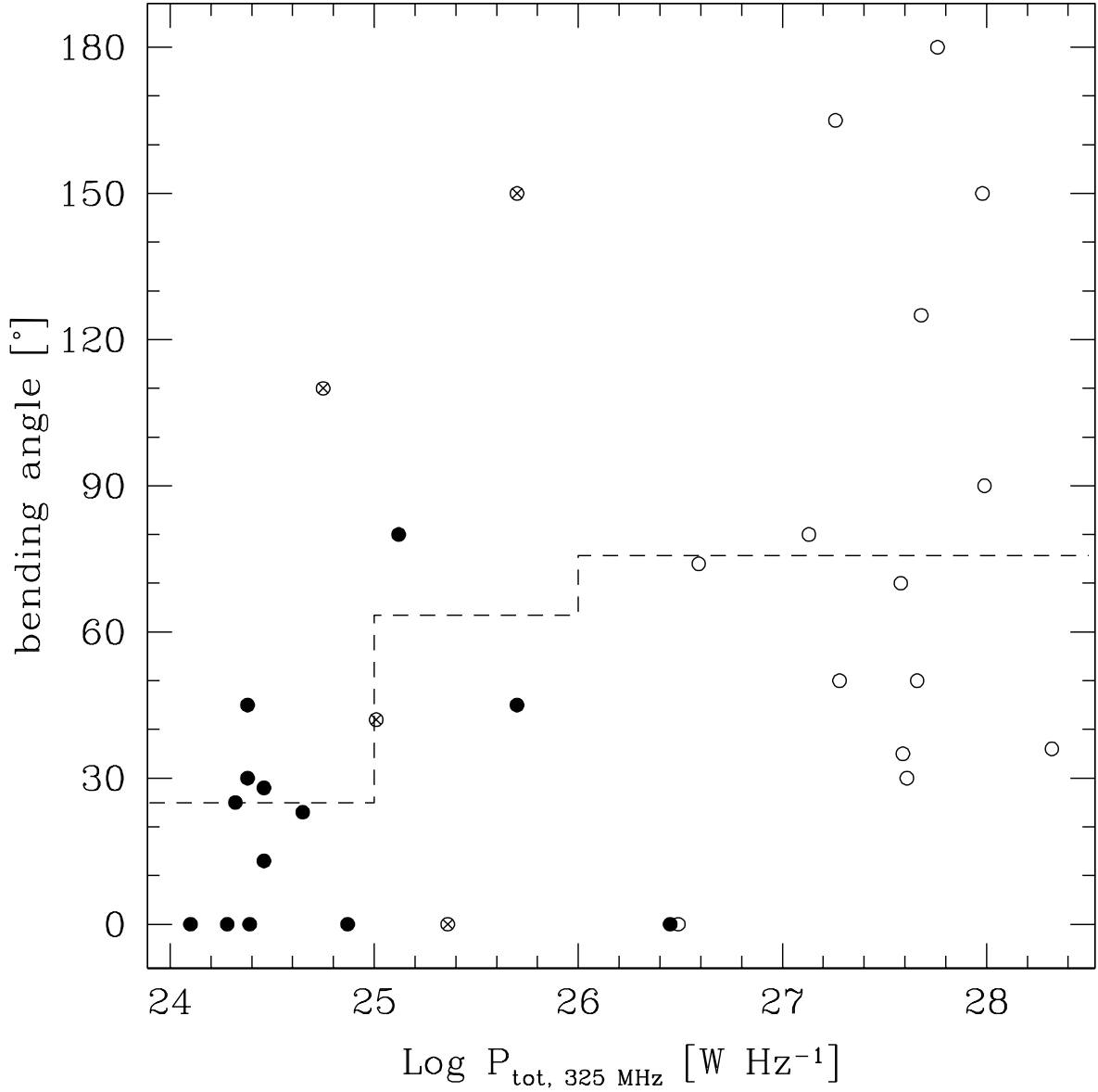}
\figcaption{Bending angle vs. total radio power at low frequency (325
  MHz). The diagram includes both BL Lacs from the present work
  (filled symbols) and the
  1 Jy sample (empty symbols); the crossed symbols are the four
  objects common to both samples (1418+546, 1514$-$241, 1652+398, and
  1807+698); the dashed line corresponds to the average bending angle
  in the three luminosity bins: $24 < \mathrm{Log}P <25$, $25 <
  \mathrm{Log}P <26$, and $\mathrm{Log}P > 26$.
\label{bending}}
\end{figure}

It is also interesting to discuss the difference in orientation with
respect to more powerful objects by considering the amount of bending
in the jets. In particular, large bending is suggestive of small
angles, while straight jets are more common in the presence of a
larger angle between the jet axis and the line-of-sight. In
Fig. \ref{histo} we present the bending angle distribution in the 1 Jy
sample\footnote{The bending angles for the objects in the 1 Jy sample
have been derived by us from maps on parsec and kpc scale published in
several works, e.g. \citet{cas99,cas02,rec01,fom00,she98}} and for the
present sample, excluding those belonging to the 1 Jy sample. The
histogram provides a comparison of the two populations, which
indicates that high power BL Lacs show larger distortions than weak
ones. The significance of the difference between the two distributions
is quite strong (Kolmogorov-Smirnov of 1.42, with a probability that
the two distribution are intrinsically similar $\la 0.03$). It is
tempting to speculate that this may be ascribed to a different jet
orientation, with smaller viewing angles for more powerful and
twisting sources. However, it is also possible to give a different
interpretation for this occurrence of large bending, if we speculate
that FR II have intrinsically more bent jets than FR I and that the 1
Jy sample has a parent population in which FR II radio galaxies are a
significant fraction. Figure \ref{bending} shows the bending angle
versus radio power for all the BL Lacs with available measurements of
the bending (from the present work or the 1 Jy sample): there is a
weak trend of larger bending in more powerful sources, with a
correlation coefficient of $r=0.42$. In particular, the lack of large
bending in low power sources is remarkable.

\begin{figure}
\plotone{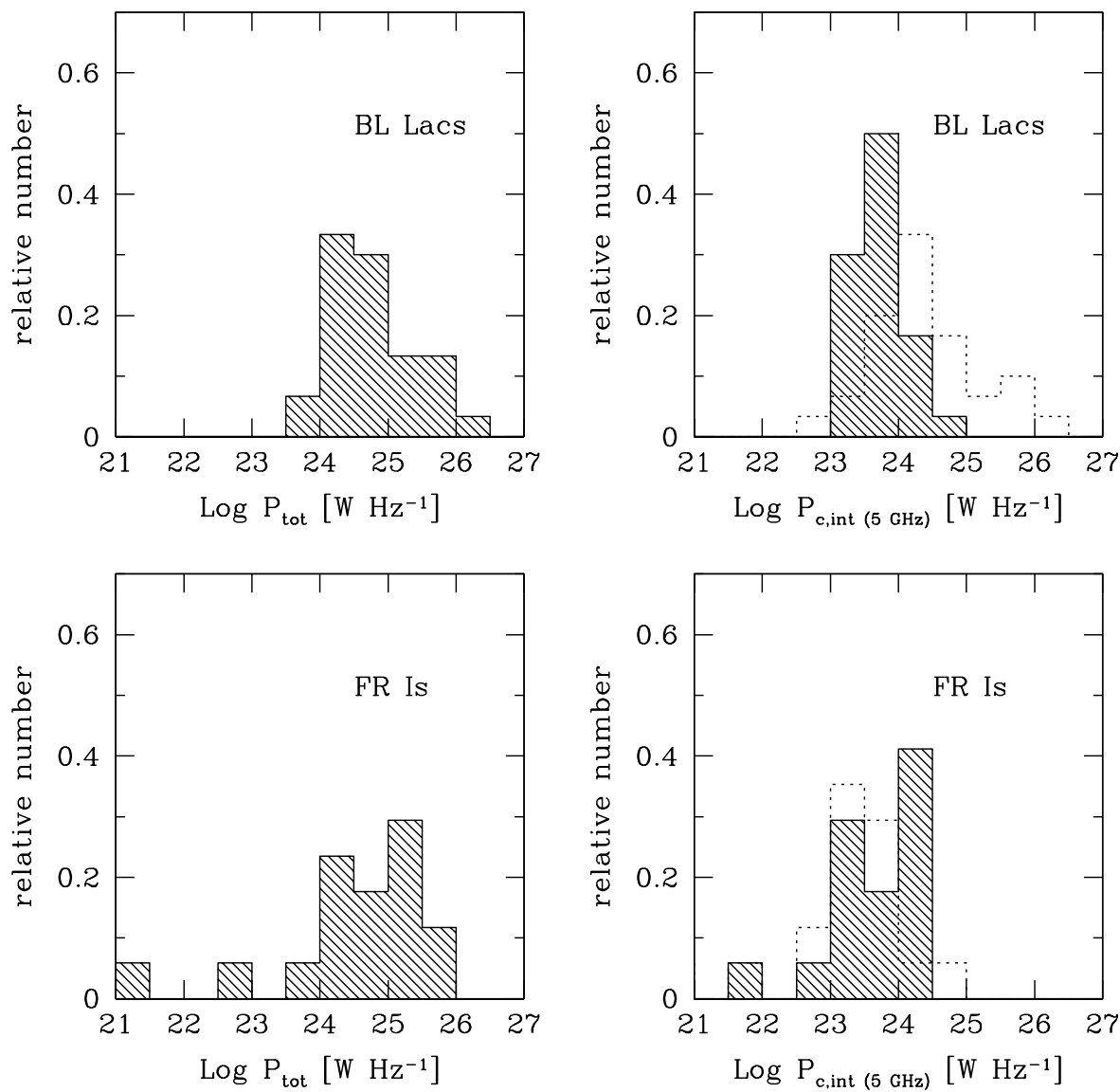}
\figcaption{Distribution of total and intrinsic core power for objects
  in the present sample and FRI and LPC radio galaxies in a sample of
  radio galaxies \citep{gio01}. The dashed histograms overlayed to the
  intrinsic core power show the distribution of the observed values.
\label{fig9}}
\end{figure}

Whilst FR II sources may contribute to the parent population of the 1
Jy sample \citep[see also][]{rec01}, FR I radio galaxies are the best
candidate to be the unbeamed counterpart of the objects in the present
sample.  From our estimate of the jet velocity and orientation, we can
derive the intrinsic core radio power: $P_{\rm c,i} = P_{\rm c,o}
\times \delta^{-(2+\alpha)}$. In Fig. \ref{fig9} we present the
distribution of the core and low frequency total radio power for the
present sample and the sample of FR I and low power compact radio
galaxies studied by \citet{gio01}. The two samples cover the same
range in low frequency total radio power (Fig. \ref{fig9}, left), as
expected if FR Is are the parent population of BL Lacs. We note that
the total radio power at 325 MHz should be an intrinsic source
property since the core is self-absorbed and extended lobe emission
dominates at low frequency. In the right panels, the shaded histograms
represent the intrinsic core radio power distribution, while the
overlaid dashed histograms refer to the observed power of the radio
core. Despite a significant difference in the observed values, the
distribution of the intrinsic core radio power is similar and in the
same range, confirming that the intrinsic properties of the two
populations are the same.

We conclude that in our sample the parent population is composed of FR
I radio galaxies alone and that the fraction of FR IIs in the parent
population of BL Lacs must be very small, if any. FR IIs that may be
present in the 1 Jy sample must therefore be ascribed to the very
large volume considered. In any case, their incidence may not be
negligible among the most powerful LBL, which could have different
properties.

\section{CONCLUSIONS}
\label{sec:conclusions}

In this paper we presented a new sample of nearby BL Lac objects with
no selection bias on their nuclear radio properties. For sources without
good data available in literature, we present new VLA and VLBA
images. Many sources exhibit a resolved radio morphology as expected
if BL Lacs are FR I radio galaxies oriented at small angles. A few
objects show a WAT or HT morphology suggesting that they are radio
galaxies belonging to cluster of galaxies. Most of the arcsecond core
flux density is present in our VLBA images; this result implies that
we have not missed considerable sub-arcsecond structure.

On the kiloparsec scale we have both symmetric and one-sided sources
suggesting that kiloparsec scale jets may have different properties,
being either non-relativistic or still mildly relativistic in
different sources. This suggests that the decrease in jet velocity is
related to the ISM properties which can vary among galaxies. More
detailed studies with the EVLA or the proposed New Mexico Array are
necessary to better investigate this possibility.

Confirming previous results by \citet{rec03}, parsec and kiloparsec
scale jets are oriented at the same PA in a large fraction of
HBLs. Given the relative numbers of HBLs and LBLs in the sample, the
low number of distorted structures and the core dominance argument,
HBL sources show less distortion and therefore are expected to be
oriented at larger angles than the LBL sources. This is confirmed by
the core dominance argument, the jet/counterjet ratio, and the
Synchrotron Self Compton model, under the assumption that all jets
possess the same Lorentz factor $\Gamma = 5$. By contrast, if we allow
both $\Gamma$ and $\theta$ to vary, we derive similar orientation
($\langle \theta_{\rm LBL} \rangle = 15^\circ \pm 7^\circ$ and
$\langle \theta_{\rm HBL} \rangle = 20^\circ \pm 4^\circ$) and a
difference in velocity; interestingly, LBL would have larger bulk
velocities (up to $\Gamma \la 7$) than HBL, including TeV sources
($\langle \Gamma \rangle \sim 3$). In both cases, the Doppler factor
of BL Lacs is considerably smaller in the parsec scale radio jets than
in the $\gamma-$ray emitting region.

In any case, we estimate that most sources ($\sim 80\%$) in our sample
are oriented at an angle to the line-of-sight larger than
$10^\circ$. The derived range in orientation correspond to a possible
range of Doppler factors. From these values, we compute the relative
intrinsic core radio power and show that intrinsic core radio powers
are similar in BL Lacs and FR I radio galaxies and low power radio
galaxies of the same range of total radio power. No FR II is allowed
as the misaligned counterpart of a low redshift BL Lac.

\begin{acknowledgments} 
 
We thank the referee Dr. Eric Perlman for many useful comments and
suggestions which improved this work. We thank also D. Dallacasa and
A. Treves for a critical reading of the paper and valuable
suggestions.  This research has made use of the NASA/IPAC
Extragalactic Database (NED) which is operated by the Jet Propulsion
Laboratory, Caltech, under contract with NASA and of NASA's
Astrophysics Data System (ADS) Bibliographic Services.  MERLIN is a
National Facility operated by the University of Manchester at Jodrell
Bank Observatory on behalf of PPARC. This material is based upon work
supported by the Italian Ministry for University and Research (MIUR)
under grant COFIN 2003-02-7534.

\end{acknowledgments}

\clearpage

\begin{deluxetable}{rllrrrcr}
\tabletypesize{\footnotesize}
\tablecaption{Objects in the Sample \label{table1}}
\tablehead{ 
\colhead{Num.} & \colhead{Name (IAU)} & \colhead{Other Name} &
\colhead{$z$} & \colhead{RA} & \colhead{Dec} & \colhead{Class} &
\colhead{Sample} }
\startdata
 1 & 0145+138   &         & 0.125 & 01 48 29.7 & +14 02 18   & H &
Slew \\
 2 & 0229+200   &         & 0.140 & 02 32 48.4 & +20 17 16   & H &
HEAO-A3 \\
 3 & 0347$-$121 &         & 0.188 & 03 49 23.2 & $-$11 59 27 & H &
HEAO-A3 \\
 4 & 0350$-$371 &         & 0.165 & 03 51 53.8 & $-$37 03 46 & H &
EMSS \\
 5 & 0521$-$365 &         & 0.055 & 05 22 58.0 & $-$36 27 31 & L &
HEAO-A3 \\
 6 & 0548$-$322 &         & 0.069 & 05 50 40.8 & $-$32 16 18 & H &
HEAO-A2 \\
 7 & 0706+591   &         & 0.125 & 07 10 30.0 & +59 08 20   & H &
HEAO-A3 \\
 8 & 0806+524   &         & 0.137 & 08 09 49.1 & +52 18 59   & H &
Slew \\
 9 & 0829+046   &         & 0.180 & 08 31 48.9 & +04 29 39   & L &
HEAO-A3 \\
10 & 0927+500   &         & 0.188 & 09 30 37.6 & +49 50 26   & H &
Slew \\
11 & 1101+384   & Mkn 421 & 0.031 & 11 04 27.3 & +38 12 32   & H &
HEAO-A3 \\
12 & 1133+704   & Mkn 180 & 0.046 & 11 36 26.4 & +70 09 27   & H &
HEAO-A3 \\
13 & 1212+078   &         & 0.136 & 12 15 10.9 & +07 32 04   & H &
Slew \\
14 & 1215+303   &         & 0.130 & 12 17 52.1 & +30 07 01   & H &
Slew \\
15 & 1218+304   &         & 0.182 & 12 21 21.9 & +30 10 37   & H &
HEAO-A2 \\
16 & 1229+643   &         & 0.164 & 12 31 31.4 & +64 14 18   & H &
EMSS \\
17 & 1255+244   &         & 0.141 & 12 57 31.9 & +24 12 40   & H &
Slew \\
18 & 1418+546   & OQ 530  & 0.152 & 14 19 46.6 & +54 23 15   & L & PG \\
19 & 1426+428   &         & 0.129 & 14 28 32.6 & +42 40 21   & H &
HEAO-A3 \\
20 & 1440+122   &         & 0.162 & 14 42 48.2 & +12 00 40   & H &
Slew \\
21 & 1514$-$241 & AP Lib  & 0.049 & 15 17 41.8 & $-$24 22 19 & H & 1
Jy \\
22 & 1652+398   & Mkn 501 & 0.034 & 16 53 52.2 & +39 45 37   & H & 1
Jy \\
23 & 1728+502   &I\,Zw 187& 0.055 & 17 28 18.6 & +50 13 10   & H &
HEAO-A3 \\
24 & 1807+698   & 3C 371  & 0.051 & 18 06 50.6 & +69 49 28   & L & 1
Jy \\
25 & 1959+650   &         & 0.048 & 19 59 59.8 & +65 08 55   & H &
HEAO-A3 \\
26 & 2200+420   & BL Lac  & 0.070 & 22 02 43.3 & +42 16 40   & L & 1
Jy \\
27 & 2201+044   &         & 0.027 & 22 04 17.6 & +04 40 02   & L &
HEAO-A3 \\
28 & 2254+074   &         & 0.190 & 22 57 17.3 & +07 43 12   & L & 1
Jy \\
29 & 2344+514   &         & 0.044 & 23 47 04.8 & +51 42 18   & H &
Slew \\
30 & 2356$-$309 &         & 0.165 & 23 59 07.8 & $-$30 37 40 & H &
HEAO-A3 \\
\enddata
\tablecomments{Col. 7: H -- High frequency peaked BL Lac, L -- Low
  frequency peaked BL Lac. Col. 8: 1 Jy \citep{sti91}; HEAO-A2
  \citep{pic82}; HEAO-A3 \citep{rem99}; EMSS \citep{mor91}; Slew
  \citep{sch93,per96}}
\end{deluxetable}

\begin{deluxetable}{lccl}
\tablecaption{Summary of 1.4 GHz VLA Observations \label{table2}}
\tablehead{
\colhead{Date} & \colhead{Conf.} & \colhead{hrs.} & \colhead{Targets}
}
\startdata
2002 Feb 22 &A& 5 &0706+591, 0806+524, 0829+046, 0927+500, 1133+704, \\
            & &   &1212+078, 1218+304, 1229+643, 1255+244, 1426+428, \\
            & &   &1728+502, 1959+650 \\
2002 May 03 &A& 2.5 &0145+138, 0229+200, 0347$-$121, 0350$-$371, 2254+074, \\
            & &     &2344+514, 2356$-$309 \\
2002 Oct 08 &C& 4.5 &0145+138, 0229+200, 0347$-$121, 1728+502, 1807+698, \\
            & &     & 1959+650, 2254+074, 2344+514, 2356$-$309 \\
\enddata
\end{deluxetable}

\begin{deluxetable}{lcl}
\tablecaption{Summary of 5 GHz VLBA Observations \label{table3}}
\tablehead{
\colhead{Date} & \colhead{hrs} & \colhead{Targets}
}
\startdata
2002 Feb 17 & 6 & 0145+138, 0229+200, 0347$-$121, 0350$-$371, 2344+514, \\
            &   & 2356$-$309 \\
2002 Feb 18 & 4 & 0521$-$365, 0548$-$322, 0706+591, 0806+524 \\
2002 Feb 19 & 5 & 1212+078, 1218+304, 1229+643, 1255+244, 1440+122 \\
\enddata
\end{deluxetable}

\begin{deluxetable}{rlrrrrrr}
\tablecaption{Image Parameters\label{table4}}
\tablehead{ 
\colhead{} & \colhead{} & \multicolumn{2}{c}{VLA (A conf.)} &
\multicolumn{2}{c}{VLA (C conf.)} & \multicolumn{2}{c}{VLBA} \\
\colhead{Num.} & \colhead{Name} & \colhead{noise} & \colhead{peak} &
\colhead{noise} & \colhead{peak} & \colhead{noise} & \colhead{peak}
}

\startdata
 1 & 0145+138 & 0.15    &   2.8 & 0.24    &   11.2 & 0.55  &  3.5\\
 2 & 0229+200 & 0.30	&  40.4	& 0.35	  &   52.9 & 0.49  & 15.7	\\
 3 &0347$-$121& 0.30	&   8.7	& 0.50	  &   10.4 & 0.53  &  7.7	\\
 4 &0350$-$371& 0.30	&  22.3	& \nodata & \nodata& 0.72  & 11.9	\\
 5 &0521$-$365& \nodata	& \nodata & \nodata & \nodata& 5.50  & 1034.9 \\
 6 &0548$-$322& \nodata	& \nodata & \nodata & \nodata& 0.90  & 33.3 	\\
 7 & 0706+591 & 0.15	&  62.6	& \nodata & \nodata& 0.38  & 33.2	\\
 8 & 0806+524 & 0.20	& 157.6	& \nodata & \nodata& 0.38  & 78.9	\\
 9 & 0829+046 & 0.25	& 736.2	& \nodata & \nodata& \nodata & \nodata\\
10 & 0927+500 & 0.11	&  19.9	& \nodata & \nodata& \nodata & \nodata\\
12 & 1133+704 & 0.18	& 111.6	& \nodata & \nodata& \nodata & \nodata\\
13 & 1212+078 & 0.12	&  83.2	& \nodata & \nodata& 0.36  & 33.1	\\
15 & 1218+304 & 0.20	&  65.6	& \nodata & \nodata& 0.41  & 44.1	\\
16 & 1229+643 & 0.12	&  53.0	& \nodata & \nodata& 0.61  & 19.6	\\
17 & 1255+244 & 0.10	&   6.5	& \nodata & \nodata& 0.55  &  2.9	\\
19 & 1426+428 & 0.10	&  31.9	& \nodata & \nodata& \nodata & \nodata\\
20 & 1440+122 & \nodata	& \nodata & \nodata & \nodata& 0.47  & 17.1	\\
23 & 1728+502 & 0.30	& 201.0	& 0.30	  &  196.7 & \nodata & \nodata\\
24 & 1807+698 & \nodata	& \nodata & 0.60  & 1348.8 & \nodata & \nodata\\
25 & 1959+650 & 0.25	& 198.5	& 0.24	  &  241.2 & \nodata & \nodata\\
28 & 2254+074 & 0.30	& 407.0	& 0.90	  &  490.4 & \nodata & \nodata\\
29 & 2344+514 & 0.30	& 195.8	& 0.40	  &  220.2 & 0.39  & 89.7	\\
30 &2356$-$309& 0.25	&  39.2	& 0.30	  &   40.2 & 0.66  & 14.2	\\
\enddata
\tablecomments{Noise levels and peak flux densities for Fig. 2 and 3
  (Col. 3, 4), for Fig. 4 (Col. 5, 6), and for Fig. 6 (Col. 7, 8). All
  units are mJy/beam.}
\end{deluxetable}

\begin{deluxetable}{rllrrrrrlc}
\tablecaption{Flux Densities \label{table5}}
\tablehead{ 
\colhead{Num.} & \colhead{Name} & \colhead{$z$} & \colhead{$S_{325
    {\rm MHz}}$}& \colhead{$S_{\rm t, 1.4 GHz}$} & \colhead{$S_{\rm
    c, 1.4 GHz}$} & \colhead{c/tot} & \colhead{$S_{\rm c, 5 GHz}$} &
    \colhead{ref.} & \colhead{morph.} \\
\colhead{} & \colhead{(IAU)} &  & \colhead{(mJy)} & \colhead{(mJy)} &
    \colhead{(mJy)} & & \colhead{(mJy)} 
}
\startdata
 1 & 0145+138 & 0.125 & 42    $\pm$ 8\tablenotemark{a}  &  35.8 & 3                    & 0.08 & 2.1  & (1)  & j   \\
 2 & 0229+200 & 0.140 & 110   $\pm$ 22\tablenotemark{a} &  94.2 & 42                   & 0.44 & 45   & (1)  & x   \\
 3 &0347$-$121& 0.188 & 30    $\pm$ 6\tablenotemark{a}  &  25.7 & 9                    & 0.35 & 8.4  & (1)  & j   \\
 4 &0350$-$371& 0.165 & 46    $\pm$ 9\tablenotemark{a}  &  40.1 & 23                   & 0.57 & 17   & (2)  & c   \\
 5 &0521$-$365& 0.055 &	42920 $\pm$ 480\tablenotemark{b}& 17620 & 3124\tablenotemark{d}& 0.18 & 2581 & (3)  & x   \\
 6 &0548$-$322& 0.069 & 1463  $\pm$ 293\tablenotemark{c}&   505 & 76\tablenotemark{e}  & 0.15 & 68   & (4)  & w   \\
 7 & 0706+591 & 0.125 & 371   $\pm$ 3.8			&   160 & 65                   & 0.41 & 80   & (5)  & j   \\
 8 & 0806+524 & 0.137 & 165   $\pm$ 4.1			& 183   & 160                  & 0.87 & 172  & (1)  & c   \\
 9 & 0829+046 & 0.180 & 671   $\pm$ 31\tablenotemark{c} & 1257  & 643\tablenotemark{d} & 0.51 & 1230 & (6)  & w   \\
10 & 0927+500 & 0.188 & 12    $\pm$ 3.7			& 22.3  & 21                   & 0.94 & 18   & (1)  & c   \\
11 & 1101+384 & 0.031 & 1374  $\pm$ 3.3			& 889   & 548\tablenotemark{e} & 0.62 & 640  & (7)  & j   \\
12 & 1133+704 & 0.046 & 529   $\pm$ 5.2			& 337   & 115                  & 0.34 & 125  & (8)  & h   \\
13 & 1212+078 & 0.136 & 175   $\pm$ 35\tablenotemark{a} & 150   & 85                   & 0.57 & 91   & (1)  & c   \\
14 & 1215+303 & 0.130 & 1175  $\pm$ 2.9			& 593   & 355\tablenotemark{d} & 0.60 & 445  & (9) & h   \\
15 & 1218+304 & 0.182 & 82    $\pm$ 2.7			& 71.3  & 67                   & 0.94 & 56   & (9) & c   \\
16 & 1229+643 & 0.164 & 72    $\pm$ 3.7			& 63    & 55                   & 0.87 & 42   & (10) & c   \\
17 & 1255+244 & 0.141 & 19    $\pm$ 4\tablenotemark{a}  & 16.6  & 6.5                  & 0.39 & 6.9  & (1)  & c   \\
18 & 1418+546 & 0.152 & 962   $\pm$ 3.4			& 835   & 1058\tablenotemark{d}& 1.27 & 1189 & (11) & x   \\
19 & 1426+428 & 0.129 & 64    $\pm$ 3.1			& 61.3  & 32                   & 0.52 & 22   & (8)  & h   \\
20 & 1440+122 & 0.162 & 80    $\pm$ 16\tablenotemark{a} & 68.5  & 60                   & 0.87 & 41   & (1)  & c   \\
21 &1514$-$241& 0.049 & 1792  $\pm$ 43\tablenotemark{c} & 2177  & 2562\tablenotemark{d}& 1.18 & 1480 & (12) & x   \\
22 & 1652+398 & 0.034 & 1936  $\pm$ 4.7			& 1630  & 1376\tablenotemark{d}& 0.84 & 1250 & (7)  & j   \\
23 & 1728+502 & 0.055 & 317   $\pm$ 4.2			& 232   & 200                  & 0.86 & 134  & (8)  & j   \\
24 & 1807+698 & 0.051 & 4015  $\pm$ 4.4			& 2129  & 1350\tablenotemark{d}& 0.63 & 1507 & (13) & h   \\
25 & 1959+650 & 0.048 & 252   $\pm$ 3.5			& 260   & 200                  & 0.77 & 252  & (1)  & j   \\
26 & 2200+420 & 0.070 & 1819  $\pm$ 4.1			& 6250  & 3310\tablenotemark{d}& 0.53 & 4857 & (9) & x   \\
27 & 2201+044 & 0.027 &	1222  $\pm$ 58\tablenotemark{c} & 835   & 179\tablenotemark{d} & 0.21 & 168  & (8)  & j   \\
28 & 2254+074 & 0.190 & 405   $\pm$ 25\tablenotemark{c} & 404   & 405                  & 1.00 & 1216 & (9) & c   \\
29 & 2344+514 & 0.044 &	569   $\pm$ 4.2			& 410   & 196                  & 0.48 & 212  & (1)  & x   \\
30 &2356$-$309& 0.165 & 75    $\pm$ 15\tablenotemark{a} & 66.2  & 39                   & 0.59 & 29   & (5)  & x   \\
\enddata
\tablecomments{Col. 4: total source flux density at 325 MHz from the
  WENSS, except as noted: (a) extrapolated from the 1.4 GHz flux density
  (Col. 5) with $\alpha = 0.11$, (b) interpolated from NED data
  between 80 and 8800 MHz, (c) Texas Survey \citep{dou96}. Col 5:
  total source flux density at 1.4 GHz from the NVSS. Col 6: arcsecond
  core flux density at 1.4 GHz from VLA A configuration data or (d)
  \citet{ant85}, (e) \citet{lau93}. Col. 7: ratio between core and
  total flux density at 1.4 GHz. Col 8: arcsecond core flux density at
  5 GHz, with references as given in Col. 9. Col. 9: (1)
  \citet{per96}, (2) \citet{ver93}, (3) interpolated from data between
  1.4 GHz and 15 GHz, (4) \citet{rei99}, (5) NED, (6) MERLIN archive
  data, 
  (7) \citet{gio01}, (8) \citet{lau93}, (9)
  \citet{fos98}, (10) \citet{pad95}, (11) \citet{mur93}, (12)
  \citet{mor93}, (13) \citet{cas99}. Col. 10: (c) unresolved; (j)
  core+jet(s), (h) core+halo, (x) extended, complex structure, (w)
  wide angle tail}
\end{deluxetable}

\begin{deluxetable}{rllrrcrrrrl}
\tabletypesize{\footnotesize}
\tablecaption{VLBA data @5 GHz \label{table6}}
\tablehead{ 
  \colhead{Num.} & \colhead{Name} & \colhead{$z$} & \colhead{$S_{t}$} &
  \colhead{$S_{c}$} & \colhead{$S_{\rm t, mas}/S_{\rm c,''}$} & \colhead{$B_{j}$} & \colhead{PA$_{\rm jet}$} &
  \colhead{$R$} & \colhead{$\beta \cos \theta$}
  & \colhead{ref.} \\ 
  &   &   & \colhead{(mJy)} & \colhead{(mJy)} & & \colhead{(mJy/beam)} &
  \colhead{($^\circ$)} & \colhead{($\ge$)} & \colhead{($\ge$)}
}

\startdata
 1 & 0145+138 & 0.125 & 3.3	& 3.2     & 1.61 & \nodata & \nodata & \nodata & \nodata & (1) \\
 2 & 0229+200 & 0.140 & 18.2	& 17.0    & 0.40 & \nodata & \nodata & \nodata & \nodata & (1)  \\
 3 &0347$-$121& 0.188 & 8.5	& 8.1     & 1.01 & \nodata & \nodata & \nodata & \nodata & (1) \\
 4 &0350$-$371& 0.165 & 16.9	& 14.0    & 0.99 & 2.40    & 46      &   8 & \nodata & (1) \\
 5 &0521$-$365& 0.055 & 1747	& 995.0   & 0.68 & 142.5   & $-$44   & 119 & 0.74 & (1) \\
 6 &0548$-$322& 0.069 & 41.7	& 36.3    & 0.61 & \nodata & \nodata & \nodata & \nodata & (1) \\
 7 & 0706+591 & 0.125 & 42.0	& 33.8    & 0.53 & 3       & $-$158  &  25 & \nodata & (1) \\
 8 & 0806+524 & 0.137 & 136.8	& 66.0    & 0.80 & 40      & 20      & 333 & 0.82 & (1) \\
 9 & 0829+046 & 0.180 & 740	& 440.0   & 0.60 & 90      & 65      &  90 & \nodata & (2) \\
10 & 0927+500 & 0.188 & \nodata & \nodata & \nodata & \nodata & \nodata & \nodata & \nodata \\
11 & 1101+384 & 0.031 & 400	& 200     & 0.63 & 10      & $-$32   & 100 & 0.73 & (3) \\
12 & 1133+704 & 0.046 & 137.9	& 82.7    & 1.10 & 9.5     & 105     &  16 & \nodata & (4) \\
13 & 1212+078 & 0.136 & 50.0	& 34.5    & 0.53 & 3.0     & 92      &  23 & \nodata & (1) \\
14 & 1215+303 & 0.130 & 372.0	& 276.0   & 0.84 & 9.4     & 140     & 157 & 0.77 & (3) \\
15 & 1218+304 & 0.182 & 56.9	& 39.8    & 1.02 & 3.2     & 73      &  25 & \nodata & (1) \\
16 & 1229+643 & 0.164 & 35.7	& 26.9    & 0.85 & 5.0     & $-$43   &  28 & \nodata & (1) \\
17 & 1255+244 & 0.141 & 2.8	& 2.8     & 0.40 & \nodata & \nodata & \nodata & \nodata & (1) \\
18 & 1418+546 & 0.152 & 1090	& 209.0   & 0.92 & 40      & 120     &  40 & \nodata & (5) \\
19 & 1426+428 & 0.129 & 21.2	& 19.1    & 0.96 & 2.1     & 20      &   3 & \nodata & (4) \\
20 & 1440+122 & 0.162 & 18.5	& 17.2    & 0.45 & \nodata & \nodata & \nodata & \nodata & (1) \\
21 &1514$-$241& 0.049 & 2278	& 1948    & 1.49 & 253     & 161     & 90  & \nodata & (6)\\
22 & 1652+398 & 0.034 & 1118	& 491     & 0.89 & 100     & 150     & 400 & 0.83 & (7) \\
23 & 1728+502 & 0.055 & 171	& 110.0   & 1.28 & 18.0    & $-55$   & 200 & 0.79 & (3) \\
24 & 1807+698 & 0.051 & 780	& 560.0   & 0.52 & 50      & $-$102  &  38 & \nodata & (2) \\
25 & 1959+650 & 0.048 & 220	& 181.7   & 0.87 & 1.6     & $-$5    &   8 & \nodata & (8) \\
26 & 2200+420 & 0.070 & 2164	& 1148    & 0.45 & 300     & $-170$  & 350 & 0.82 & (9) \\
27 & 2201+044 & 0.027 & 170	& 131.3   & 1.01 & 3.0     & $-42$   &   3 & \nodata & (4) \\
28 & 2254+074 & 0.190 & 350	& 280.0   & 0.29 & 38.4    & $-120$  &  43 & \nodata & (2) \\
29 & 2344+514 & 0.044 & 116.7	& 70.0    & 0.55 & 3.0     & 142     &  25 & \nodata & (1) \\
30 &2356$-$309& 0.165 & 21.9    & 21.9    & 0.76 & \nodata & \nodata & \nodata & \nodata & (1) \\

\enddata
\tablecomments{Col. 4: Correlated VLBI flux density at 5 GHz. Col. 5:
  VLBI core flux density. Col. 6: Ratio between total VLBI flux
  density and arcsecond core at 5 GHz (Col. 8 in \ref{table4}. Col. 7:
  Jet brightness near the core. Col. 8: Jet P.A. Col. 9: Lower limit
  for the jet/counterjet ratio. Col. 10: Estimated lower limit for
  $\beta \cos \theta$. Col. 11: References -- (1) present work, (2)
  \citet{fey00}, (3) Giroletti et al., in prep., (4) \citet{kol96},
  (5) \citet{cas02}, (6) \citet{fom00}, (7) \citet{gir04}, (8)
  \citet{rec03}, (9) \citet{gab03} }
\end{deluxetable}

\begin{deluxetable}{lrrr}
\tablecaption{Limits from the Synchrotron Self Compton model \label{table7}}
\tablehead{ 
  \colhead{Name} & \colhead{$\delta_{\rm min}$} &
  \colhead{$\Gamma_{\rm min}$} & \colhead{$\theta_{\rm max}$} \\
  \colhead{} & \colhead{} & \colhead{} & \colhead{($^\circ$)}
}
\startdata
0521$-$365& 2.2 & 1.3 & 27  \\
0829+046 &  4.6 & 2.4 & 13  \\
1101+384 &  2.4 & 1.4 & 25  \\
1215+303 &  2.6 & 1.5 & 22  \\
1418+546 &  2.2 & 1.3 & 26  \\
1514$-$241& 9.3 & 4.7 & 6   \\
1652+398 &  2.5 & 1.4 & 24  \\
1807+698 &  5.3 & 2.7 & 11  \\
2200+420 &  6.7 & 3.4 & 9   \\
2201+044 &  1.1 & 1.1 & 65  \\
2254+074 &  4.0 & 2.1 & 15  \\
\enddata
\tablecomments{Col. 2: Minimum Doppler factor from SSC (we only report
  sources where $\delta_{\rm min} > 1$). Col. 3: Minimum Lorentz
  factor $\Gamma$. Col. 4: Largest possible orientation angle.}
\end{deluxetable}

\begin{deluxetable}{llcccc}
\tablecaption{Final intrinsic parameters with $\Gamma = 5$\label{table8}}
\tablehead{
  \colhead{Num.} & \colhead{Name} & \colhead{${\rm Log}P_{\rm t}$} & \colhead{$\Delta
    \theta_5$} & \colhead{$\Delta \delta_5$} & \colhead{$\Delta P_{\rm
      c,i}$} \\
  \colhead{} & \colhead{} & \colhead{W ${\rm Hz}^{-1}$} & \colhead{($^{\circ}$)} &
  \colhead{} & \colhead{W ${\rm Hz}^{-1}$}
}
\startdata
 1 & 0145+138 & 24.17 & 41 -- 61 &  0.4 --  0.8 & 23.10 -- 23.70 \\
 2 & 0229+200 & 24.69 & 19 -- 29 &  1.4 --  2.8 & 23.42 -- 24.02 \\
 3 &0347$-$121& 24.39 & 23 -- 35 &  1.0 --  2.0 & 23.24 -- 23.84 \\
 4 &0350$-$371& 24.46 & 21 -- 32 &  1.2 --  2.4 & 23.28 -- 23.88 \\
 5 &0521$-$365& 26.45 & 21 -- 27 &  1.5 --  2.3 & 24.51 -- 24.86 \\
 6 &0548$-$322& 25.19 & 32 -- 48 &  0.6 --  1.2 & 23.73 -- 24.33 \\
 7 & 0706+591 & 25.12 & 21 -- 32 &  1.2 --  2.4 & 23.69 -- 24.29 \\
 8 & 0806+524 & 24.85 & 12 -- 21 &  2.4 --  4.7 & 23.52 -- 24.12 \\
 9 & 0829+046 & 25.70 &  3 -- 13 &  4.6 --  9.1 & 24.05 -- 24.65 \\
10 & 0927+500 & 23.99 & 15 -- 24 &  2.0 --  3.9 & 22.99 -- 23.59 \\
11 & 1101+384 & 24.46 & 19 -- 25 &  1.8 --  2.7 & 23.28 -- 23.63 \\
12 & 1133+704 & 24.39 & 25 -- 37 &  0.9 --  1.8 & 23.23 -- 23.83 \\
13 & 1212+078 & 24.87 & 16 -- 26 &  1.7 --  3.4 & 23.53 -- 24.13 \\
14 & 1215+303 & 25.65 & 14 -- 22 &  2.1 --  4.0 & 24.02 -- 24.57 \\
15 & 1218+304 & 24.80 & 15 -- 24 &  1.9 --  3.7 & 23.49 -- 24.09 \\
16 & 1229+643 & 24.65 & 17 -- 27 &  1.6 --  3.2 & 23.40 -- 24.00 \\
17 & 1255+244 & 23.93 & 24 -- 37 &  0.9 --  1.9 & 22.95 -- 23.55 \\
18 & 1418+546 & 25.71 &  7 -- 15 &  3.7 --  7.5 & 24.05 -- 24.65 \\
19 & 1426+428 & 24.38 & 22 -- 33 &  1.1 --  2.2 & 23.23 -- 23.83 \\
20 & 1440+122 & 24.68 & 17 -- 27 &  1.5 --  3.1 & 23.42 -- 24.02 \\
21 &1514$-$241& 24.97 &  0 --  6 &  7.7 --  9.9 & 22.90 -- 23.11 \\
22 & 1652+398 & 24.69 & 16 -- 24 &  1.9 --  3.5 & 23.42 -- 23.93 \\
23 & 1728+502 & 24.32 & 21 -- 32 &  1.2 --  2.4 & 23.19 -- 23.79 \\
24 & 1807+698 & 25.36 &  0 -- 11 &  5.3 --  9.9 & 22.94 -- 23.48 \\
25 & 1959+650 & 24.10 & 16 -- 26 &  1.7 --  3.3 & 23.06 -- 23.66 \\
26 & 2200+420 & 25.29 &  3 --  9 &  6.4 --  9.2 & 23.80 -- 24.11 \\
27 & 2201+044 & 24.28 & 30 -- 44 &  0.7 --  1.3 & 23.17 -- 23.77 \\
28 & 2254+074 & 25.53 &  0 -- 11 &  5.4 --  9.9 & 24.02 -- 24.54 \\
29 & 2344+514 & 24.38 & 21 -- 32 &  1.2 --  2.3 & 23.23 -- 23.83 \\
30 &2356$-$309& 24.67 & 19 -- 30 &  1.3 --  2.7 & 23.41 -- 24.01 \\
\enddata
\tablecomments{Col. 3: Logarithm of total radio power at 325 MHz. Col. 4:
  range of possible jet orientation angle. Col. 5: Doppler factor
  range. Col. 6: Logarithm of intrinsic core radio power at 5 GHz. }
\end{deluxetable}

\clearpage

\end{document}